\documentclass[aps,prl,a4paper,twocolumn,superscriptaddress,citeautoscript,footinbib,floatfix]{revtex4-2}
\usepackage{color}
\usepackage[english]{babel}
\usepackage{lmodern}
\usepackage[T1]{fontenc}
\usepackage{amsmath}
\usepackage{units}
\usepackage{grffile}
\usepackage{color}
\usepackage[bottom]{footmisc}
\usepackage{booktabs}
\usepackage{dcolumn}
\usepackage{graphicx}
\usepackage{lipsum}
\usepackage{changes}
\graphicspath{{figs/}}
\begin{document}
%
\title{On the Spatial Locality of Electronic Correlations in LiFeAs
}
\author{Minjae Kim}
\email{garix.minjae.kim@gmail.com}
\affiliation{Department of Physics and Astronomy, Rutgers University, Piscataway, New Jersey 08854, USA}
\affiliation{Department of Chemistry, Pohang University of Science and Technology (POSTECH), Pohang 37673, Korea}
\author{Hu Miao}
\affiliation{Materials Science and Technology Division, Oak Ridge National Laboratory, Oak Ridge, Tennessee 37831, USA}
\author{Sangkook Choi}
\affiliation{Condensed Matter Physics and Materials Science Department, Brookhaven National Laboratory, Upton, New York 11973, USA}
\author{Manuel Zingl}
\affiliation{Center for Computational Quantum Physics, Flatiron Institute, 162 5th Avenue, New York, NY 10010, USA}
\author{Antoine Georges}
\affiliation{Coll\`ege de France, 11 place Marcelin Berthelot, 75005 Paris, France}
\affiliation{Center for Computational Quantum Physics, Flatiron Institute, 162 5th Avenue, New York, NY 10010, USA}
\affiliation{Centre de Physique Th\'eorique, \'Ecole Polytechnique, CNRS, Universit\'e Paris-Saclay, 91128 Palaiseau, France}
\affiliation{Department of Quantum Matter Physics, University of Geneva, 24 Quai Ernest-Ansermet, 1211 Geneva 4, Switzerland}
\author{Gabriel Kotliar}
\affiliation{Department of Physics and Astronomy, Rutgers University, Piscataway, New Jersey 08854, USA}
\affiliation{Condensed Matter Physics and Materials Science Department, Brookhaven National Laboratory, Upton, New York 11973, USA}
\date{\today}
\begin{abstract}
We address the question of the degree of spatial non-locality of the self energy in the iron-based superconductors,
a subject which is receiving considerable attention. Using LiFeAs as a prototypical example, we extract the self energy
from angular-resolved photoemission spectroscopy (ARPES) data. We use two distinct electronic structure references:
density functional theory in the local density approximation and linearized quasiparticle
self consistent GW (LQSGW).
We find that with the LQSGW reference, spatially local dynamical correlations provide a consistent description of the experimental data, and account for some surprising aspects of the data such as the substantial out of plane
dispersion of the electron Fermi surface having dominant $xz/yz$ character.
Hence, correlations effects can be separated into static non-local contributions well described by LQSGW
and dynamical local contributions.
Hall effect and resistivity data are shown to be consistent with this description.
\end{abstract}
\maketitle

{\it Introduction.}
The origin  of superconductivity in the iron pnictides and chalcogenides is an 
outstanding open problem in condensed matter physics\cite{kamihara2006iron}.
Two opposite points of view have been presented.
In the first one, superconductivity originates from the exchange of spatially non-local antiferromagnetic (AFM) spin fluctuations~\cite{mazin2008unconventional,kuroki2008unconventional,chubukov2008magnetism} and
non-local correlations are also essential in the normal state~\cite{Zantout_Non_locality,fink2019evidence,dai2015antiferromagnetic,fanfarillo2012unconventional}.
The second one posits a more local pairing due to Hund's coupling \cite{umezawa2012unconventional,hoshino2015superconductivity,miao2018universal,lee2018pairing,coleman2020triplet},
which in turns requires a rather local picture of the normal state.

Answering this question requires a proper understanding of the degree of spatial locality of electronic correlations in the normal state. This has been addressed previously by a comparison of theoretical calculations to experiments.
Some results favor the
local picture~\cite{yin2011kinetic,miao2012isotropic,lee2012orbital,werner2012satellites,yin2014spin,miao2015observation,
miao2016orbital,semon2017validity,tomczak2012many}
while others  support the non local
view~\cite{ferber2012lda,Zantout_Non_locality,fink2019evidence,ortenzi2009fermi,bhattacharyya2020non}.

Here, we take a different approach and address this question by a direct examination of experimental data
from angle-resolved photoemission spectroscopy (ARPES), following the   approach which was successful for Sr$_2$RuO$_4$~\cite{tamai2019high}.
We consider LiFeAs \cite{tapp2008lifeas}, a prototypical iron-based superconductor which is free from magnetic and nematic instabilities
and which has been intensively studied for more than a decade~\cite{dai2015antiferromagnetic,yin2014spin,yin2011kinetic},
%
and use the experimentally measured quasiparticle dispersions for the different Fermi surface (FS) sheets to
determine the self-energy and assess its degree of spatial locality.
%

Our results offer a solution to the local vs. non-local conundrum\cite{mertz2018self}. 
We find that the electronic self-energy can be separated, to a good approximation, into a non-local part which is
frequency independent, and a dynamical (frequency-dependent) part which is spatially local to a good approximation.
The non-local part can be incorporated in the reference Hamiltonian with respect to which the dynamical self-energy
is defined, and we show that the quasiparticle GW approximation \cite{tomczak2015qsgw,kutepov2017linearized,kutepov2012electronic} provides a good starting point to that effect.
These findings are in line with previous work by Tomczak et al.~\cite{tomczak2012many,tomczak2015qsgw},
but we emphasize that our conclusions are established directly from {\it experimental observations}, once the proper  reference Hamiltonian to define self energies is used. This finding rationalizes the success of
dynamical mean field theory (DMFT)~\cite{DMFT_rmp_1996,kotliar2006electronic} for these materials~\cite{qazilbash2009electronic,de2014selective,yin2011kinetic},
and emphasizes GW+DMFT as a method of choice in this context~\cite{tomczak2012many,tomczak2015qsgw,choi_comdmft:_2019}.


{\it Method.}
%
Ignoring photoemission matrix elements, extrinsic and surface effects,
we relate the measured photoemission spectra to the spectral function associated with the
one particle Green's function:
\begin{eqnarray} \label{eq:Greens_Function}
&G(k,\omega)=&[\omega\cdot\textbf{I}-H(k)-\Sigma(k,\omega)]^{-1}_{m\sigma,m'\sigma'}
\end{eqnarray}
In this expression, $H(k)$ is a reference Hamiltonian matrix expressed in a localised basis of orbitals $m,\sigma$
($\sigma$ is the spin index),
$\omega$ is the frequency, and $k$ is the wavevector in the Brillouin zone.
$\Sigma(k,\omega)$ is the self-energy matrix for the given reference Hamiltonian $H(k)$.
The chemical potential is included in $H(k)$.

We consider two different choices for the  reference Hamiltonian $H(k)$. The first is the Kohn-Sham Hamiltonian
obtained from density-functional theory
in the local density approximation (DFT-LDA) using the Wien2k software package~\cite{blaha2001wien2k,blaha2020wien2k}.
The second is the quasiparticle Hamiltonian obtained from the linearized quasiparticle self-consistent GW method
(LQSGW) using the FlapwMBPT code~\cite{kutepov2012electronic,kutepov2017linearized}.
For the localised basis set ($m\sigma$), we calculate maximally localized Wannier functions \cite{souza2001maximally,marzari1997maximally}
in a wide energy window including Fe($d$) and As($p$) orbitals, using the Wannier90 \cite{mostofi2008wannier90}, Wien2Wannier \cite{kunes_wien2wannier:_2010} and ComDMFT \cite{choi_comdmft:_2019} packages.
(See the Supplemental Material (SM) ~\footnote{
See Supplemental Material (SM)
for (i) information on the construction of $H(k)$ with maximally localized Wannier function (MLWF) for DFT-LDA and LQSGW,
(ii) details of the microscopic calculations include SOC,
(iii) $k_{z}$ dependent electron band dispersions of $H(k)$,
(iv) details of the method for the extraction of the self-energy including error bars,
(v) comparison of the quality of the present LQSGW+non-local $\Sigma$ fit to the published
ARPES data, showing that the fitting of the hole pockets leads to
a good descriptions of the published data for $k_{z}$=0.00 \cite{miao2016orbital}
and for other values of $k_{z}$.\cite{wang2015topological},
(vi) self-energy depending on the assigned $k_{z}$ value for electron pockets and corresponding Fermi surface volumes for the Luttinger's theorem,
(vii) frequency dependency of the dynamical self-energy, and
(viii) analysis of the transport data of Ref.\cite{rullier2012multiorbital}
and discussions of the extracted scattering rate from the transport data.
}.)
We take the spin-orbit coupling (SOC) to be local and present only on iron atoms (see SM \cite{Note1}).

\begin{table}[t]
\caption{The net Fermi surface volumes, V$^{\text{electron}}_{\text{FS,total}}$-V$^{\text{hole}}_{\text{FS,total}}$,
and Fermi surface volumes of each sheet (electrons/unit cell)
in (a) the LDA, (b) the LDA+non-local $\Sigma$ ansatz, ($k_{z}$=0.00 for fitting of hole bands and $k_{z}$=0.35 for fitting of electron bands),
(c) the LDA+local $\Sigma$ ansatz, ($k_{z}$=0.00 for fitting of hole bands and $k_{z}$=0.35 for fitting of electron bands),
(d) the LQSGW,
(e) the LQSGW+non-local $\Sigma$ ansatz, ($k_{z}$=0.00 for fitting of hole bands and $k_{z}$=0.55 for fitting of electron bands).
0.02-0.03 (electrons/unit cell) in the net Fermi surface volume is the numerical uncertainty.
}
\begin{tabular}{|| m{3.1cm} || m{0.68cm} | m{0.68cm} | m{0.68cm} | m{0.68cm} | m{0.68cm} | m{0.8cm} ||}
\hline
 ~ & $\alpha$' & $\alpha$ & $\beta$ & $\gamma$ & $\delta$ & Net \\
\hline
 LDA & 0.01 & 0.14 & 0.33 & 0.18 & 0.28 & -0.02 \\
\hline
 LDA+non-local $\Sigma$ & 0.00 & 0.08 & 0.37 & 0.23 & 0.39 & +0.17 \\
\hline
 LDA+local $\Sigma$ & 0.00 & 0.06 & 0.36 & 0.19 & 0.35 & +0.12 \\
\hline
 LQSGW & 0.00 & 0.08 & 0.35 & 0.20 & 0.21 & -0.03 \\
\hline
 LQSGW+non-local $\Sigma$ & 0.00 & 0.05 & 0.36 & 0.20 & 0.26 & +0.04 \\
\hline
\end{tabular}
\label{table:FS_volumes}
\end{table}

\begin{figure}[t]
\includegraphics[width=\columnwidth]{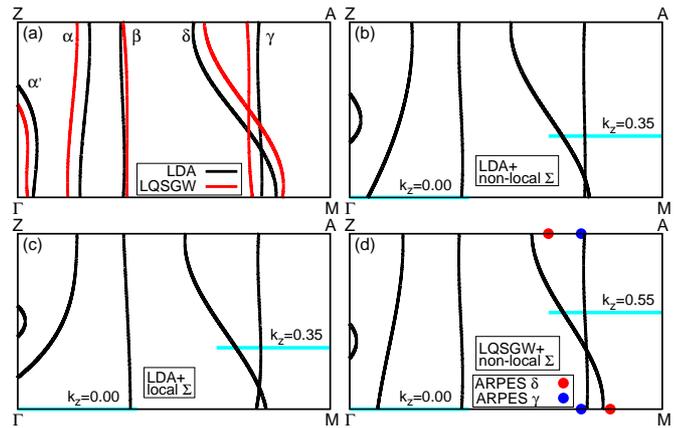}
\caption{ (a) Fermi surfaces of LDA (black) and LQSGW (red) in the $\Gamma$-$M$-$A$-$Z$ plane
(b) Same as (a) for the LDA+non-local $\Sigma$ ansatz with $k_{z}$=0.00 for the hole bands fit
and $k_{z}$=0.35 for the electron bands fit.
(c) Same as (a) for the LDA+local $\Sigma$ ansatz with $k_{z}$=0.00 for the hole bands fit
and $k_{z}$=0.35 for the electron bands fit.
(d) Same as (a) for the LQSGW+non-local $\Sigma$ ansatz with $k_{z}$=0.00 for the hole bands fit
and $k_{z}$=0.55 for the electron bands fit.
Red and blue dots are the $\delta$ and $\gamma$ Fermi surfaces
measured with ARPES in Ref.\cite{brouet2016arpes}.
The ARPES data for fitting (b-d) is taken from Ref.\cite{miao2016orbital,Note2}.
See Table~\ref{table:FS_volumes} for Fermi surface volumes.
\label{fig:FS_3D}
}
\end{figure}

We first discuss the electronic structure associated with $H(k)$, i.e. in the absence of the self-energy.
In Fig.~\ref{fig:FS_3D}(a), we compare the FS of DFT-LDA to that of LQSGW.
The LQSGW FS clearly displays a significant shrinking of the $xz/yz$ dominated hole/electron pockets,
$\alpha'$, $\alpha$, and $\delta$ sheets in comparison to LDA, as pointed
out in previous work~\cite{tomczak2015qsgw,tomczak2012many,Zantout_Non_locality}.
This is because non-local electronic interactions are more prominently taken into account in the LQSGW,
resulting in a repulsion of the bands between $\alpha$ ($\alpha'$) and $\delta$.
The shrinking of these FS pockets from LDA to LQSGW is also apparent from Table~\ref{table:FS_volumes}, in which
we compare the volumes of the different FS sheets between the two methods. The net difference between
all electron and hole FS volumes  is also indicated and, for both methods, adds up to zero within error bars
as required by the Luttinger's theorem.

The procedure for extracting the self-energy from ARPES data follows Ref.~\cite{tamai2019high} for Sr$_{2}$RuO$_{4}$. From a theoretical viewpoint, the dispersions of the different branches
of quasiparticles are the solutions of $\det[\omega-H(k)-\mathrm{Re}\Sigma(k,\omega)]=0$
(neglecting the lifetime effects associated with $\mathrm{Im}\Sigma$).
We use the
measured positions of the maximum of the momentum distribution curves (MDC) associated with several quasiparticle bands, for
a given binding energy $\omega$, as an input to this equation which is then solved by a numerical root-finding procedure
for the real part of the self-energy (for details of the procedure, see SM~\cite{Note1}).

\begin{figure}[t]
\includegraphics[width=\columnwidth]{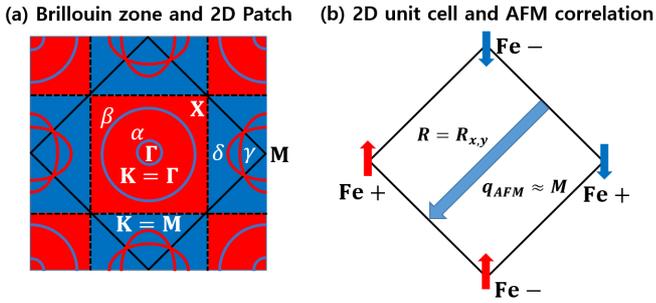}
\caption{(a) Patching of the Brillouin zone for the non-local $\Sigma$ ansatz of LiFeAs.
The solid line delimits the principal Brillouin zone
(two irons in a unit cell), and the dashed lines
indicate the patching
used in the non-local $\Sigma$ ansatz. 
The patch centered on K=$\Gamma$ (resp. K=$M$) is colored in red (resp. blue).
Schematic Fermi surfaces are represented by colored
solid lines. Hole pockets are in blue, $\alpha$ (inner) and $\beta$ (outer). Electron pockets are in red, $\delta$ (outer)
and $\gamma$ (inner).
(b) Two-dimensional unit cell and the momentum $q_{AFM}\approx M$ associated with AFM correlations~\cite{qureshi2012inelastic}. The AF-correlated Fe moments are schematized by the blue and red arrows,
with Fe+ and Fe- denoting the two Fe atoms in the unit cell. 
\label{fig:BZ}
}
\end{figure}

To facilitate the determination of $\Sigma$, we restrict its functional form as follows.
We assume that, in the local orbital basis, it is independent of the out-of-plane
momentum $k_z$ and that the  off-diagonal (inter-orbital) matrix elements are
absorbed into the renormalization of the SOC\cite{kim2018spin,linden2020imaginary,horvat2017spin}.
Two different {\it ans\"{a}tze} are made for the in-plane momentum dependence.
(i) The self-energy components are simply assumed to be independent of momentum - we refer
to this as the `local $\Sigma$ ansatz'.
(ii) The Brillouin zone is divided into two patches, centered around the
$\Gamma$- and $M$-points, respectively, as illustrated on Fig.~\ref{fig:BZ}, and
a more flexible momentum dependence is allowed which is piecewise constant in each patch.
We refer to this ansatz as the 'non-local $\Sigma$ ansatz'
It corresponds to a two-site dynamical cluster approximation
(DCA) which is a cluster extension of the DMFT \cite{maier2005quantum}.
These two ans\"{a}tze thus read (see SM for details~\cite{Note1}):
\begin{eqnarray} \label{eq:Self_Energy}
\textrm{local $\Sigma$~ansatz:}~\Sigma_{m}(k,\omega)&=&\Sigma_{m}(\omega)~ \\
\nonumber
\textrm{non-local $\Sigma$~ansatz:}~\Sigma_{m}(k,\omega)&=&\Sigma_{m}(\Gamma,\omega)
\,\,\,\mathrm{if}\,\,\,k\in \Gamma \nonumber\\
&=&\Sigma_{m}(M,\omega)\,\,\,\mathrm{if}\,\,\,k\in M\,\,\,
\end{eqnarray}
The components of the self-energy within the non-local $\Sigma$ ansatz are obtained by
fitting the experimental hole bands at K=$\Gamma$ and electron bands at K=$M$ separately.
We also note that this ansatz is physically motivated by the AFM wave-vector of spin fluctuations
and corresponding Brillouin zone folding (Fig.~\ref{fig:BZ}(b))~\cite{qureshi2012inelastic}.
We emphasize that these {\it ans\"{a}tze} are made for the components of the
self-energy expressed in the basis of local orbitals. The transformation to the quasiparticle (band) basis
is momentum dependent and leads to significant momentum dependence of the self-energy in
that basis even if a DMFT ansatz is made (see also Ref.~\cite{tamai2019high}).

%
%
The
 assignment of $k_z$ from  ARPES has     uncertainties ~\cite{damascelli2004probing}. In our case,
experiment constraints
 $k_{z}$ around the electron pockets to be in the interval  $[0.3,0.7]$  while
for the hole pockets, there is little uncertainty that the data arise from $k_{z}=0$ \footnote{Hu Miao, unpublished},
(See SM\cite{Note1}).
For the electron pockets, we considered two different ways to infer $k_z$.
(i) The first is to require that the Fermi surface volume satisfies Luttinger's theorem,
as obtained by a full Brillouin zone integration
and assuming that the self-energy does not depend on $k_z$.
As it turns out, for electron pockets, a unique value of $k_z\simeq 0.55$ satisfies this constraint for both ans\"{a}tze.
%
(ii) The second one determines $k_z$ by requesting that the resulting self energy is as local
as possible.  This leads to $k_{z}$=0.35 for the  LDA+non-local $\Sigma$ ansatz and $k_{z}$=0.55 for the LQSGW+non-local $\Sigma$ ansatz (see SM\cite{Note1}).
Note that in that case, Luttinger's theorem is violated within the LDA+non-local $\Sigma$ ansatz, while the value
$k_z=0.55$ ensures both Luttinger's theorem and maximal locality when using the LQSGW reference.

\begin{figure}[t]
\includegraphics[width=\columnwidth]{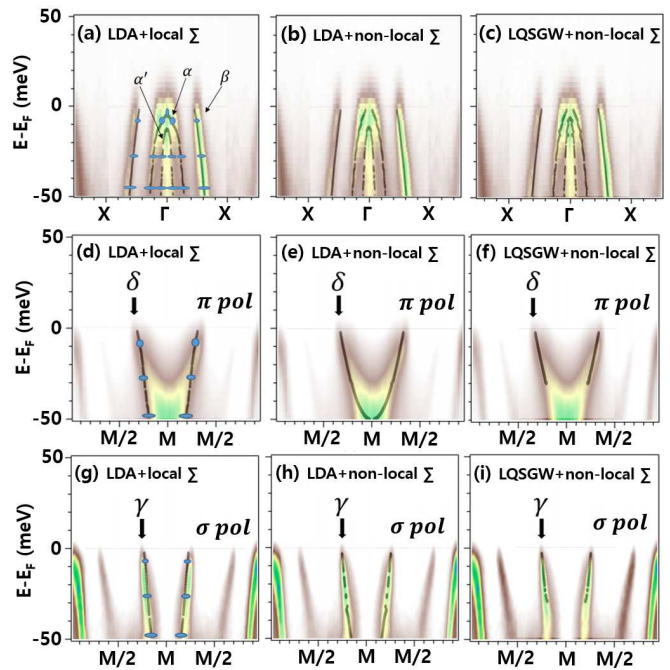}
\caption{Comparison between the ARPES data of Ref.~\cite{miao2016orbital,Note2} for LiFeAs (color intensity map)
with different polarizations ($\pi$ and $\sigma$) and the quasiparticle dispersions obtained
with the different ans\"{a}tze discussed in the text.
The results of  LDA+local $\Sigma$, LDA+non-local $\Sigma$,and LQSGW+non-local $\Sigma$ $ans\"{a}tze$
are shown in panels  (a, d and g),    (b, e and h),   and  (c, f and i), respectively.
The hole pocket data are taken at $k_{z}$=0.00 and the
electron pockets data are taken at a value of $k_{z}$ within the range of [0.3, 0.7]~\cite{miao2016orbital,Note2}.
The blue colored ellipses are theoretical error estimates arising from the width of the MDC
peak and the uncertainty in $k_{z}$.
For the electron bands, $k_{z}$=0.55 has been used in the fit
using LQSGW as a reference, while $k_{z}$=0.35 when using LDA as a reference (see main text).
\label{fig:Fit}
}
\end{figure}

{\it Results.}
Our main results are summarized in
Figs.~\ref{fig:FS_3D}(b-d),
\ref{fig:Fit} and Tables~\ref{table:FS_volumes} and \ref{table:Linear}.
The full frequency dependence of the self-energies extracted from the procedure described above is displayed
on Fig.S5 in the supplemental material (SM)\cite{Note1}.
All results were obtained using
the ARPES data of Ref\cite{miao2016orbital,Note2}, also displayed
on Fig.~\ref{fig:Fit}~\footnote{The ARPES data has been measured at 20 K
which is slightly above the superconducting transition temperature (18 K) of LiFeAs.\cite{tapp2008lifeas}}.
%
%
The low-energy behaviour of the fitted self-energies is characterized by
the zero-frequency (static) values $\Sigma_m(0)$, as well as the quasiparticle weights
$Z_m=\left[1-\frac{\partial\Sigma_m}{\partial\omega}|_{\omega=0}\right]^{-1}$, displayed
in Table~\ref{table:Linear}.
Comparing the values obtained within the non-local $\Sigma$ ansatz for the $\Gamma$- and $M$- BZ patches, we see
that, when starting from LDA, the static components of the self-energy are spatially local 
to a good approximation for
the $xz/yz$ orbitals, while a higher degree of momentum-dependence holds for the $xy$ orbital.
%
The quasiparticle weight associated with the $xy$ orbital is found to be weakly momentum dependent, while
stronger momentum dependence is found for the $xz/yz$ orbital.
This strong momentum dependence of the dynamical self-energy of the $xz/yz$ orbitals has been
discussed in Refs.~\cite{fink2019evidence,bhattacharyya2020non,Zantout_Non_locality,miao2015observation} in relation to
the strong coupling of the quasiparticles of the $xz/yz$ driven $\alpha$ and $\alpha'$
hole-like FS sheets to the existing AFM correlation in LiFeAs~\cite{qureshi2012inelastic}.
Indeed, these FS sheets are close to the AFM zone boundary.
The values of the quasiparticle weights obtained here, (0.15 ($\Gamma$) 0.12 ($M$) for $xy$
and 0.25 ($\Gamma$) 0.16 ($M$) for $xz/yz$),
are smaller than that of the computed LDA+DMFT values reported in Refs.~\cite{yin2011kinetic,miao2016orbital,lee2012orbital}
(Z$_{xy}$=0.26 and Z$_{xz/yz}$=0.34).
They are however close to the values ($0.17-0.19$) reported
by de Haas-van Alphen experiments~\cite{putzke2012haas}.

\begin{table}[t]
\caption{
Zero frequency self-energy ($\Sigma_{m}(\mathrm{K},0)$)
and quasiparticle residue (Z$_{m}$(K)) extracted from ARPES
data of LiFeAs\cite{miao2016orbital,Note2},
with the LDA+non-local $\Sigma$ ansatz, the LQSGW+non-local $\Sigma$ ansatz,
and the LDA+local $\Sigma$ ansatz.
We use $k_{z}$=0.00 for K=$\Gamma$ (hole sheets) for both the LDA and the LQSGW references, $k_{z}$=0.35 for K=$M$ (electron sheets) for the LDA reference, and
$k_{z}$=0.55 for K=$M$ (electron sheets), for the LQSGW reference.
Error bars (total)
are computed from the peak width of both in plane $k$ and
out of plane $k_{z}$. (See SM for the details
on the definition of the error bars\cite{Note1}.)
}
\begin{tabular}{|| c || c | c | c | c ||}
\hline
\multicolumn{5}{||c||}{LDA+non-local $\Sigma$ ansatz} \tabularnewline
\hline
 ~ & $\Sigma_{m}(\Gamma,0)$ (eV)  & Z$_{m}$($\Gamma$) & $\Sigma_{m}(M,0)$ (eV) & Z$_{m}$($M$) \tabularnewline
\hline
$xy$ & 0.029$\pm$0.025 & 0.15$\pm$0.01 & -0.130$\pm$0.062 & 0.12$\pm$0.01 \tabularnewline
\hline
$xz/yz$ & -0.083$\pm$0.040 & 0.25$\pm$0.13 & -0.113$\pm$0.026 & 0.16$\pm$0.03 \tabularnewline
\hline
\multicolumn{5}{||c||}{LDA+local $\Sigma$ ansatz} \tabularnewline
\hline
 ~ & \multicolumn{2}{|c|}{$\Sigma_{m}(0)$ (eV)}  & \multicolumn{2}{|c||}{Z$_{m}$} \tabularnewline
\hline
$xy$ & \multicolumn{2}{|c|}{0.023} & \multicolumn{2}{|c||}{0.14}\tabularnewline
\hline
$xz/yz$ & \multicolumn{2}{|c|}{-0.112} & \multicolumn{2}{|c||}{0.17} \tabularnewline
\hline
\multicolumn{5}{||c||}{LQSGW+non-local $\Sigma$ ansatz} \tabularnewline
\hline
 ~ & $\Sigma_{m}(\Gamma,0)$ (eV)  & Z$_{m}$($\Gamma$) & $\Sigma_{m}(M,0)$ (eV) & Z$_{m}$($M$) \tabularnewline
\hline
$xy$ & 0.002$\pm$0.014 & 0.21$\pm$0.01 & 0.044$\pm$0.036 & 0.18$\pm$0.01 \tabularnewline
\hline
$xz/yz$ & -0.027$\pm$0.003 & 0.38$\pm$0.01 & -0.051$\pm$0.114 & 0.30$\pm$0.04 \tabularnewline
\hline
\end{tabular}
\label{table:Linear}
\end{table}

Table~\ref{table:Linear} also displays the results obtained by using  a local ansatz for the
self-energies. As seen there, the values of the quasiparticle weights are intermediate between the values at
the $\Gamma$- and $M$-points obtained within the non-local ansatz.
%

%
%
 Fig.~\ref{fig:FS_3D}(b,c) displays how the FS is modified by  self-energy effects
when using LDA as a starting point. Table~\ref{table:FS_volumes} reports the corresponding volume of each FS sheet.
We see that both the local and non-local ansatz lead to a violation of the Luttinger theorem, when the value
$k_{z}$=0.35 is used for the fitting of electron bands.
This is mostly due to the large volume obtained for the $\delta$-sheet, which crosses the $\gamma$-sheet
at a low value of $k_{z}\approx 0.05$ leading to a too large electron-like contribution.
%
%
%
%
%
%

%
We now turn to the results of the self-energy obtained by using LQSGW for the reference Hamiltonian, using
$k_z=0.55$ in this case when fitting the electron bands around $M$.
The results in Table~\ref{table:Linear} clearly show that the fitted values of
both $\Sigma_m(0)$ and $Z_m$ are quite momentum independent (spatially local) within
the determined error bars. Some slight momentum dependence of $Z_{xz/yz}$ is found however
($\sim 0.38$ at $\Gamma$-point vs $\sim 0.30$ at $M$-point), close to the limit set by
error bars. (See Fig.S5 of the SM \cite{Note1} for the full frequency dependence of the
extracted self-energies)
Furthermore the Luttinger theorem is now well obeyed  (Table~\ref{table:FS_volumes}).
This is due in particular to the much smaller inflation of the volume
of the $\gamma$ and $\delta$ sheets by self-energy effects, in comparison to the
LDA starting point. Correspondingly, the crossing point between the
$\gamma$ and $\delta$ sheets occurs at a larger value of $k_z$ (Fig.~\ref{fig:FS_3D}(d)).
Comparing to available experimental data, we see that the LQSGW reference combined with a quasi-local self-energy
provides:
(i) a good description of the $k_{z}$ dependent
hole bands ($\alpha$', $\alpha$, and $\beta$) dispersions in comparison to  the ARPES data of
Refs~\cite{miao2016orbital,wang2015topological} (see SM~\cite{Note1} for comparison),
(ii) a good description of the $k_{z}$
dependent $\gamma$ FS in ARPES of Refs~\cite{brouet2016arpes,hajiri2012three},
(iii) and a qualitative description of the
$k_{z}$ dependent $\delta$ FS in ARPES
with correct $k_{z}$ for the crossing of the $\delta$ and $\gamma$
FSs and somewhat larger curvature
of the $\delta$ FS near the momentum of A of Ref.\cite{brouet2016arpes,Note2}
Fig.\ref{fig:FS_3D}(d) implies that for
electron bands, the overall amplitude of $k_{z}$ dependent
variation of the $\delta$ and $\gamma$ FSs
in the LQSGW+ fit is consistent
with the ARPES data of Ref.\cite{brouet2016arpes}.


%

We compare in Fig.\ref{fig:Fit}(a-c) the experimental ARPES intensity to the  fitted hole bands
of LiFeAs   using  the different starting points $H(k)$ and ansatz for $\Sigma$.
For the $xy$ dominant $\beta$ band, all schemes  compare well  with ARPES.
In contrast, we observe some differences between the different ansätze (comparable to error bars) for the
position of the top of the $\alpha $ band with dominant $xz/yz$ character.
The LDA+local $\Sigma$ ansatz leads to a lower energy than the LDA+non-local $\Sigma$ ansatz and the LQSGW+non-local $\Sigma$ ansatz.
The splitting of the states with $xz/yz$ character at the $\Gamma$ point is controlled by the
SOC and given by $\lambda\,Z_{xz/yz}$\cite{kim2018spin}.
Its experimental value is 9.5-11.4 meV\cite{borisenko2016direct,day2018influence}.
The values of $\lambda$ in LDA and LQSGW are $50$~meV and $25$~meV respectively, which
when multiplied by the extracted $Z$'s from Table~\ref{table:Linear},
indeed leads to values close to $10$~meV in both cases (see SM
for details on the effect of SOC in LiFeAs\cite{Note1}).

We now turn to the electron bands in Fig.\ref{fig:Fit} (d-i).
Along the $\Gamma-M $ direction, the $\gamma$ band has almost pure $xy$
character, and is seen in $\sigma$ polarized ARPES.
This $\gamma$ band is well described by both
the non-local ansatz and the local ansatz within both references (LDA and LQSGW), see panels
Fig.\ref{fig:Fit}(g,h, and i).
For the $xz/yz$ dominant $\delta$ band, the LDA+non-local $\Sigma$ ansatz
and the LQSGW+non-local $\Sigma$ ansatz yield quasiparticle spectra which are consistent with ARPES within error bars
as shown in Fig.\ref{fig:Fit} (e, and f).
However, differences between the fits are seen for the $xz/yz$ driven $\delta$ band
with the LDA+local $\Sigma$ ansatz having a steeper dispersion and a lower bottom than the non-local $\Sigma$ ansatz,
as seen in Fig.\ref{fig:Fit}(d, e, and f).
Also, as noted in Table \ref{table:FS_volumes} and Fig.\ref{fig:FS_3D}(b,c), this fit violates  the Luttinger's theorem.
%



In summary, our analysis demonstrates that an LQSGW reference~\cite{tomczak2015qsgw,choi_first-principles_2016,choi_comdmft:_2019}
in combination with quite local self-energies provides a description of the quasiparticle dispersions of LiFeAs in good agreement
with experiments.
The strong  dispersion along $k_{z}$ of the $\alpha$ and $\delta$ FS sheets, unique to the 111 compounds,
is also well described, although the latter is slightly overestimated.
Hence, correlation effects can be decomposed into non-local, frequency independent contributions captured by the LQSGW
and dynamical frequency-dependent contributions that are spatially local to a good approximation.
In contrast, when using LDA as a reference, the extracted self-energy
is spatially non-local and, when taken to be $k_z$-independent, leads to an overestimation of the volume of
the FS $\delta$-sheet and a corresponding violation of Luttinger theorem.
We emphasize that, in contrast to theories attributing non-locality to AFM spin fluctuations,
the non-locality in the LQSGW approach originates from the charge sector.

We finally turn to transport measurements, as reported in Ref.~\cite{rullier2012multiorbital}, and
investigate whether our LQSGW+local $\Sigma$ analysis is consistent with those data. Using the
occupancies of the different FS sheets obtained above, we use the experimental data for the resistivity
and Hall effect to obtain the scattering rates 
associated with each orbital component, as a function of temperature, under the assumption that they are spatially local.
The conclusion of this analysis (see details in SM~\cite{Note1}) is that the $xy$ orbital is found to have a larger scattering rate than
the $xz/yz$ one, and that it undergoes a clear crossover at $T\sim 150$~K between a high-$T$ incoherent regime to
a low-$T$ coherent one. This is consistent with the LQSGW+local $\Sigma$ finding that the $xy$ orbital is the more correlated one.
At low-$T$ both scattering rates are found to have a Fermi liquid $T^2$ behaviour.
As a consistency check, we also obtain satisfactory agreement with the magnetoresistance data.
Let us emphasize that, in contrast, studies emphasizing non-locality due to low-energy AFM fluctuations yield
a non-Fermi liquid scattering rate of the  $xz/yz$ orbital which is larger
than that of $xy$~\cite{fink2019evidence,Zantout_Non_locality}.


Several authors have pointed at some discrepancies between experimental data and the predictions of LDA+DMFT, which is
usually interpreted as a failure of the DMFT to take into account non-local
effects~\cite{fink2019evidence,Zantout_Non_locality,bhattacharyya2020non,ortenzi2009fermi,borisenko2010superconductivity}.
Here, based on a direct analysis of ARPES experimental data, we presented a very different picture,
consistent with the electronic structure+DMFT conceptual framework. 
We have shown that when we start from the  LQSGW reference Hamiltonian, the low energy self-energy is spatially local,
satisfies  Luttinger's theorem,  describes  available experimental data  well  and therefore is an attractive platform to study how  superconductivity emerges at lower temperatures. \cite{coleman2020triplet,miao2012isotropic,miao2018universal,umezawa2012unconventional,yin2014spin,lee2018pairing}.

\acknowledgements
We acknowledge useful discussions with Andrea Damascelli and Ryan Day (who also kindly shared their unpublished ARPES data) as
well as with Roser Valenti.
This work was supported by the DOE CMS program (MK and GK).
SC was supported by the U.S Department of Energy, Office of Science, Basic Energy Sciences as a part of the Computational Materials Science Program. For the LQSGW calculation, we used resources of the National Energy Research Scientific Computing Center (NERSC), a U.S. Department of Energy Office of Science User Facility operated under Contract No. DE-AC02-05CH11231.
HM was supported by the Laboratory Directed Research and Development Program of
Oak Ridge National Laboratory, managed by UT-Battelle, LLC, for the U.S. Department of Energy.
AG acknowledges the support of the European Research Council (ERC-319286-QMAC).
The Flatiron Institute is a division of the Simons Foundation.

\bibliography{refs_LiFeAs_Nonlocal}


\renewcommand{\thetable}{S\arabic{table}}
\renewcommand{\thefigure}{S\arabic{figure}}
\renewcommand{\thetable}{S\arabic{table}}
\renewcommand\theequation{S\arabic{equation}}
\setcounter{table}{0}
\setcounter{figure}{0}
\setcounter{equation}{0}
\renewcommand{\bibnumfmt}[1]{[S#1]}
\renewcommand{\citenumfont}[1]{S#1}

\def\cred{\color{red}}
\def\cblue{\color{blue}}

\onecolumngrid

\clearpage

\begin{center}
{\bf \large
{\it Supplemental Material:}\\
On the Spatial Locality of Electronic Correlations in LiFeAs
}

\vspace{0.2 cm}
Minjae Kim$^{1,2}$, Hu Miao$^{3}$, Sangkook Choi$^{4}$, Manuel Zingl$^{5}$, Antoine Georges$^{6,5,7,8}$, and Gabriel Kotliar$^{1,4}$
\vspace{0.1 cm}

{\small{\it
$^{1}$Department of Physics and Astronomy, Rutgers University, Piscataway, New Jersey 08854, USA \\
$^{2}$Department of Chemistry, Pohang University of Science and Technology (POSTECH), Pohang 37673, Korea \\
$^{3}$Materials Science and Technology Division, Oak Ridge National Laboratory, Oak Ridge, Tennessee 37831, USA \\
$^{4}$Condensed Matter Physics and Materials Science Department, Brookhaven National Laboratory, Upton, New York 11973, USA \\
$^{5}$Center for Computational Quantum Physics, Flatiron Institute, 162 5th Avenue, New York, NY 10010, USA \\
$^{6}$Coll\`ege de France, 11 place Marcelin Berthelot, 75005 Paris, France \\
$^{7}$Centre de Physique Th\'eorique, \'Ecole Polytechnique, CNRS, Universit\'e Paris-Saclay, 91128 Palaiseau, France \\
$^{8}$Department of Quantum Matter Physics, University of Geneva, 24 Quai Ernest-Ansermet, 1211 Geneva 4, Switzerland
}
}
\end{center}


\section{I. Construction of the reference Hamiltonian of $H^{LDA}$}

In this section, we describe details of the construction
of the reference Hamiltonian in the LDA, $H^{LDA}$.
We construct maximally localized Wannier function (MLWF) of LiFeAs for the $p-d$ model approach (Fe($d$) and As($p$))
using the Wannier90 and the Wien2Wannier packages.\cite{mostofi2008wannier90,kunevs2010wien2wannier}
We use the energy window of the interval of -6.0 eV to 3.0 eV for Fe($d$) and As($p$) bands.
We used the experimental crystal structure of LiFeAs of the Ref.\cite{pitcher2008structure}.
For the convergence of the charge density in the LDA,
we used a $k$-mesh of 10000, and checked that the charge density and the total energy
are converged with criterions of $5\times10^{-4}$ (electrons/formula unit) and 0.7 (meV/formula unit), respectively.
For the construction of MLWF, we used a $k$-mesh of $11\times11\times7$.
The local axis for the MLWF of Fe($d$) is chosen such that
(i) $z$ along $c$ of the unit cell, and
(ii) $x$ and $y$ axes toward nearest neighboring Fe atoms.
For the $p-d$ model, the spread function of the MLWF is converged as
(i) 0.985, 1.073, 1.111, and 1.333 (${\AA}^{2}$)
for $z^{2}$, $x^{2}-y^{2}$, $xy$, and $xz/yz$ orbitals of Fe,
and (ii) 3.197, and 3.301 (${\AA}^{2}$) for $p_{z}$, and $p_{x,y}$ orbitals of As.

\begin{figure}[b]
\includegraphics[width=\columnwidth]{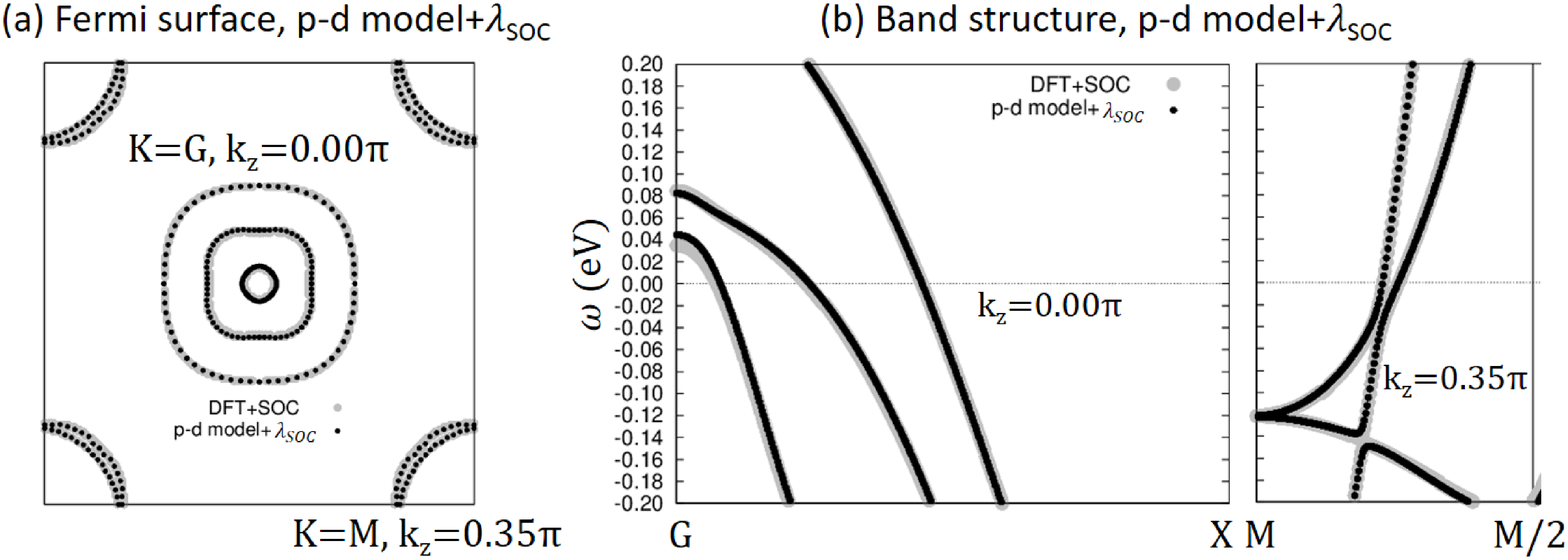}
\caption{(a) Comparison of Fermi surfaces of the LDA+SOC and
the p-d model + $\lambda_{SOC}$ (LDA+$\lambda_{SOC}$).
(b) Comparison of band dispersions of the LDA+SOC and
the p-d model + $\lambda_{SOC}$ (LDA+$\lambda_{SOC}$).
The effective SOC
of Fe($d$) is $\lambda_{SOC}$ = 50 meV.
Here we take $k_{z}$=0.00 for hole bands,
and $k_{z}$=0.35 for electron bands.
\label{fig:SOC_local}
}
\end{figure}

Fig.\ref{fig:SOC_local} presents Fermi surfaces
and low energy band structures of $p-d$ model + $\lambda_{SOC}$ (LDA+$\lambda_{SOC}$) in comparison with LDA plus SOC (LDA+SOC)
with $\lambda_{SOC}$=50 meV.
This data implies that the electronic structure of the $p-d$ model+$\lambda_{SOC}$ (LDA+$\lambda_{SOC}$) and LDA+SOC are consistent.
The factor that the local SOC is implemented only in the Fe($d$) orbital in the $p-d$ model+$\lambda_{SOC}$ (LDA+$\lambda_{SOC}$) proves that
the SOC in the LDA+SOC is fully Fe($d$) orbital driven.

\begin{figure}[t]
\includegraphics[width=\columnwidth]{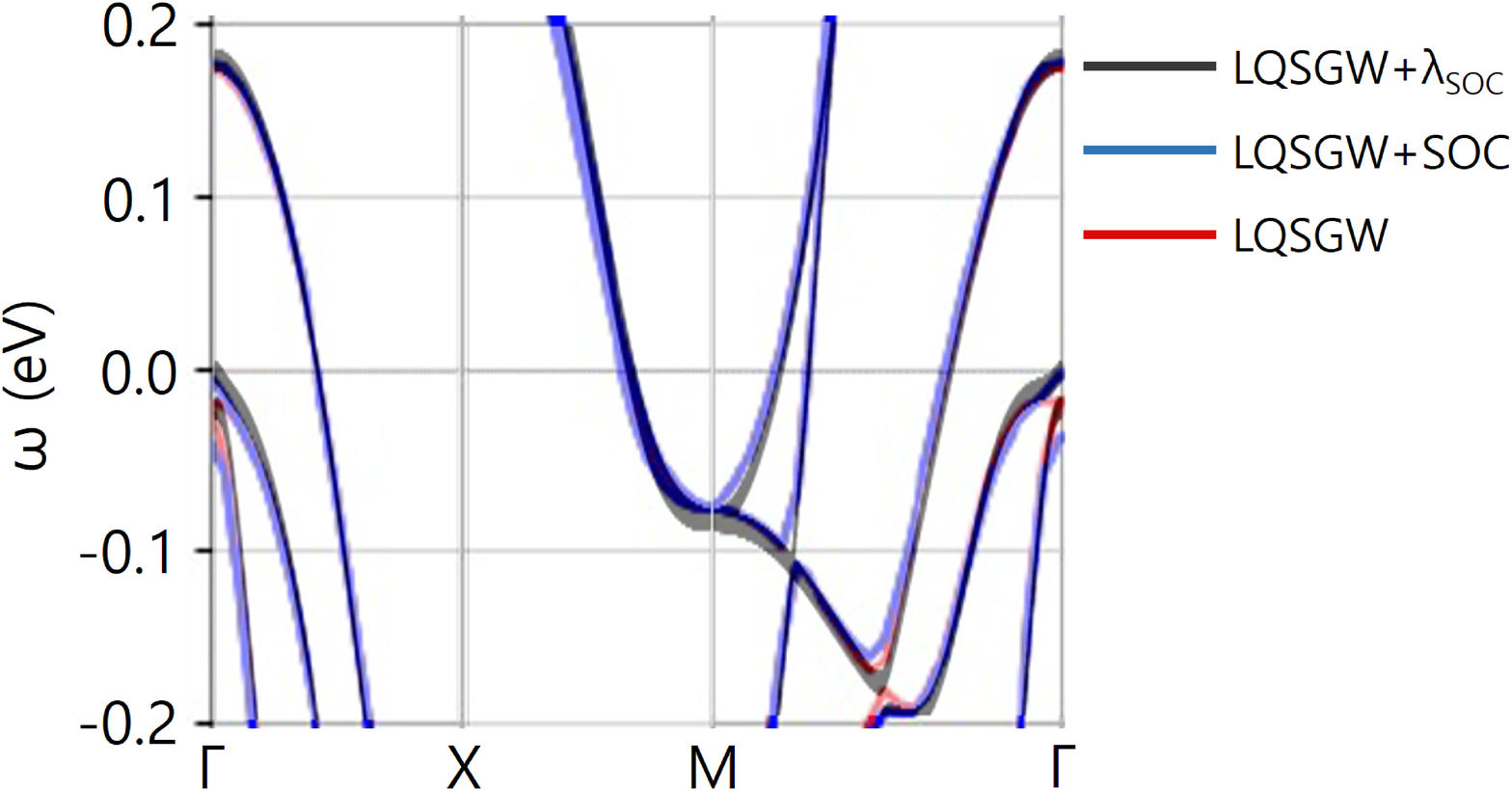}
\caption{Comparison of band dispersions of the LQSGW+$\lambda_{SOC}$ ($\lambda_{SOC}$=25 meV),
LQSGW+SOC, and LQSGW.
Here, the chemical potential is -30 meV for the Luttinger's theorem.
\label{fig:SOC_LQSGW_local}
}
\end{figure}

\section{II. Construction of the reference Hamiltonian of $H^{LQSGW}$}
In this section, we describe details of the construction
of the reference Hamiltonian in the LQSGW, $H^{LQSGW}$.
LQSGW calculation is performed by using FlapwMBPT package \cite{kutepov_electronic_2012,kutepov_linearized_2017}, which is based on full-potential linearized augmented plane wave plus local orbital method. For the crystal structure, experimental lattice constants and atomic positions \cite{tapp2008lifeas} are used. The Muffin-tin (MT) radius ($R$) is chosen in the following way: 1.74 for Li, 2.27 for Fe, and 2.16 for As in Bohr radius. Wave functions are expanded by spherical harmonics with l up to 3 for Li, 4 for Fe, and 4 for As in the MT spheres. In the interstitial region (IS), it is expanded by plane waves with the cutoff ($K_{cut}$) of $R_{Fe}\times K_{cut}$= 4.4. Product basis set is expanded by spherical harmonics with l up to 4 in the MT spheres and by planewaves with the cutoff ($G_{cut}$) of $R_{Fe}\times G_{cut}$= 7.4 in IS region.  All the unoccupied states are taken into account for both polarizability and self-energy calculation. The Brillioun zone is sampled in $6\times6\times4$ grid.

By using ComWann modules in ComDMFT package \cite{choi_comdmft:_2019} utilizing Wannier90 package \cite{mostofi2008wannier90}, 42 wannier functions are constructed: Li-p, Fe-s, Fe-p, Fe-d, As-s, As-p, and As-d orbitals.  The frozen energy window is set between -9 eV to 7 eV  and the disentanglement energy window is between -9 eV to 49 eV. Initial trial orbitals are constructed by using Muffin-tin orbitals with well-defined angular momentum characters. The local axis for the MLWF of Fe($d$) is chosen such that (i) $z$ along $c$ of the unit cell, and (ii) $x$ and $y$ axes toward nearest neighboring Fe atoms. The spread function of the MLWF is converged as (i) 0.497, 0.493, 0.524, and 0.522 (${\AA}^{2}$) for $z^{2}$, $x^{2}-y^{2}$, $xy$, and $xz/yz$ orbitals of Fe, and (ii) 1.499, and 1.724 (${\AA}^{2}$) for $p_{z}$, and $p_{x,y}$ orbitals of As.

Fig.\ref{fig:SOC_LQSGW_local} presents low energy band structures of
the LQSGW+$\lambda_{SOC}$ with local SOC for Fe($d$) (with $\lambda_{SOC}$=25 meV) in comparison with the LQSGW plus SOC (LQSGW+SOC).
The overall consistency in between the band structure of the LQSGW+SOC and the LQSGW+$\lambda_{SOC}$
implies that the local SOC in Fe($d$) is a good approximation.
The splitting of $\alpha$ and $\alpha^{,}$ bands
at $\Gamma$ is 35 meV for the LQSGW+SOC and 25 meV for the LQSGW+$\lambda_{SOC}$, respectively (See Table~\ref{table:SOC}).
In Fig.\ref{fig:SOC_LQSGW_local}, it is shown that
dispersions of $\alpha$, $\beta$, $\delta$, and $\gamma$ bands
are agreement in between the LQSGW+SOC and the LQSGW+$\lambda_{SOC}$.
These $\alpha$, $\beta$, $\delta$, and $\gamma$ bands
are taken for the extraction of the self-energy.

\begin{figure}[t!]
\includegraphics[width=15cm]{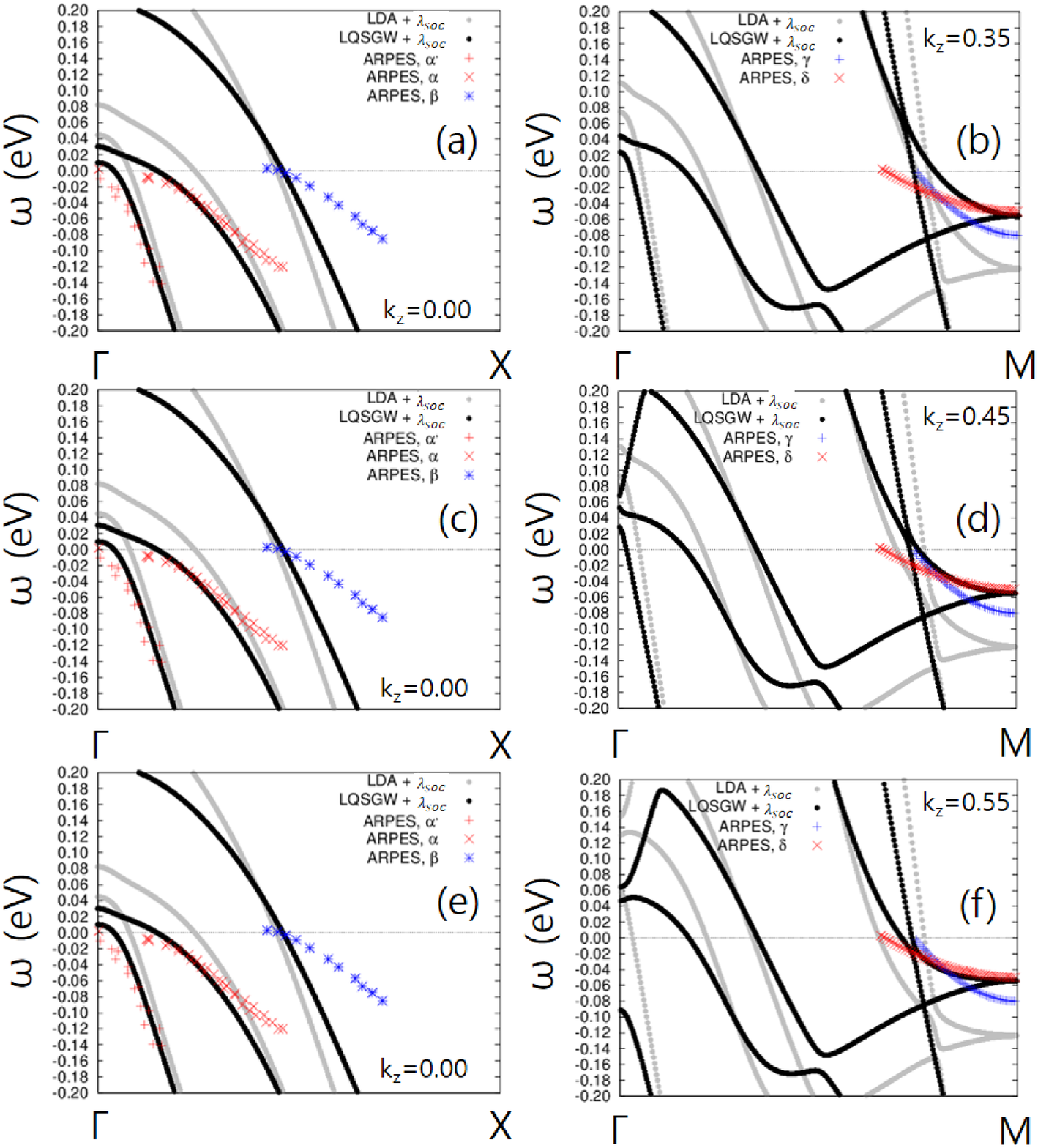}
\caption{(a) and (b) comparison
of the ARPES data to the (i) LDA reference ($\lambda_{SOC}$=50 meV)
and the (ii) LQSGW reference ($\lambda_{SOC}$=25 meV)
with $k_{z}$=0.00 for hole pockets ($\alpha$, $\alpha$', and $\beta$)
and $k_{z}$=0.35 for electron pockets ($\delta$ and $\gamma$).
(c) and (d) same as (a) and (b) with
$k_{z}$=0.00 for hole pockets ($\alpha$, $\alpha$', and $\beta$)
and $k_{z}$=0.45 for electron pockets ($\delta$ and $\gamma$).
(e) and (f) same as (a) and (b) with
$k_{z}$=0.00 for hole pockets ($\alpha$, $\alpha$', and $\beta$)
and $k_{z}$=0.55 for electron pockets ($\delta$ and $\gamma$).
\label{fig:kz}
}
\end{figure}

\section{III. ARPES versus band structures $H(k)$ of LDA and LQSGW}

In this section, we describe comparisons of bands in
between ARPES and $H(k)$ of LDA and LQSGW
to provide background information for the extraction of self-energy.

Fig.\ref{fig:kz} presents comparison of ARPES (\cite{miao2016orbital}, see the main text),
to the band structures of $H(k)$ of LDA and LQSGW.
The case of $k_{z}$=0.00 is shown for hole pockets for the extraction of self-energy.
The case of $k_{z}$=0.35-0.55 is shown for the electron pockets
to show the effect of the choice of $k_{z}$ for the extraction of the self-energy.

$LDA~reference$-For hole pockets with $k_{z}$=0.00, following factors are shown in Fig.\ref{fig:kz}(a,c, and e).
(i) There is an agreement of the size of the $xy$ driven $\beta$ Fermi surface of $H^{LDA}(k)$
to the ARPES.
(ii) There is a significant shrinking of the size of $xz/yz$ driven $\alpha$ and $\alpha'$
Fermi surfaces in the ARPES with respect to those Fermi surfaces from $H^{LDA}(k)$.
(iii) There are renormalizations of bands in the ARPES
from the $H^{LDA}(k)$ for $\alpha$, $\alpha'$, and $\beta$ bands.
For electron pockets with $k_{z}=0.35$, following factors are shown in Fig.\ref{fig:kz}(b).
(i) There is a shrinking of the $xz/yz$ driven $\delta$ band
in the ARPES with respect to the band from $H^{LDA}(k)$,
with a similar amount with respect to the case of $xz/yz$ driven $\alpha$ and $\alpha'$.
(ii) There is a small shrinking of the $xy$ driven $\gamma$ band
in the ARPES with respect to the band from $H^{LDA}(k)$.
(iii) There are renormalizations of bands in the ARPES
from the $H^{LDA}(k)$ for $\delta$ and $\gamma$ bands.
For electron pockets with $k_{z}=0.55$,
the $xy$ driven $\gamma$ band of $H^{LDA}(k)$ have a similar feature
with respect to the case with $k_{z}$=0.35, as shown in Fig.\ref{fig:kz}(b and f).
For the electron band $\delta$ with $k_{z}=0.55$,
for the size of the Fermi momentum,
there is an agreement
in between the ARPES and the $H^{LDA}(k)$ as shown in Fig.\ref{fig:kz}(f).
For $k_{z}$=0.55 for electron pockets,
there are also renormalizations of bands in ARPES
with respect to that in the $H^{LDA}(k)$ as shown in Fig.\ref{fig:kz}(f).

$LQSGW~reference$-For hole pockets with $k_{z}$=0.00, following factors are shown in Fig.\ref{fig:kz}(a,c, and e).
(i) There is an agreement of the size of the $xy$ driven $\beta$ Fermi surface of $H^{LQSGW}(k)$
with the ARPES.
(ii) There is a small shrinking of the size of $xz/yz$ driven $\alpha$ and $\alpha'$
Fermi surfaces in the ARPES with respect to those Fermi surfaces from $H^{LQSGW}(k)$.
(iii) There are renormalizations of bands in the ARPES
from the $H^{LQSGW}(k)$ for $\alpha$, $\alpha'$, and $\beta$ bands.
For electron pockets with $k_{z}=0.55$, following factors are shown in Fig.\ref{fig:kz}(f).
(i) There is a small shrinking of the $xz/yz$ driven $\delta$ band
in the ARPES with respect to the band from $H^{LQSGW}(k)$,
with a similar amount with respect to the case of $xz/yz$ driven $\alpha$ and $\alpha'$.
(ii) There is an agreement of the Fermi momentum of the $xy$ driven $\gamma$ band
in the ARPES with respect to the band from $H^{LQSGW}(k)$.
(iii) There are renormalizations of bands in the ARPES
from the $H^{LQSGW}(k)$ for $\delta$ and $\gamma$ bands.
For electron pockets with $k_{z}=0.35$, following factors are shown in Fig.\ref{fig:kz}(b).
(i) There is a significant shrinking of the $xz/yz$ driven $\delta$ band
in the ARPES with respect to the band from $H^{LQSGW}(k)$,
with a much larger amount with respect to the case of $xz/yz$ driven $\alpha$ and $\alpha'$.
(ii) There is an agreement of the Fermi momentum of the $xy$ driven $\gamma$ band
in the ARPES with respect to the band from $H^{LQSGW}(k)$.
(iii) There are renormalizations of bands in the ARPES
from the $H^{LQSGW}(k)$ for $\delta$ and $\gamma$ bands.

\section{IV. Extraction of the self-energy}

In this section, we describe the procedure for the extraction of the self-energy of LiFeAs.
For $H(k)$, we assume the reference Hamiltonian of LDA+$\lambda_{SOC}$ ($\lambda_{SOC}$=50 meV)
and LQSGW+$\lambda_{SOC}$ ($\lambda_{SOC}$=25 meV) as described in sections I and II.
%
%

In order to extract the self-energy (real part) from the ARPES data, we define
the $|d|$ of Eq.\ref{eq:method} from the Green's function which is defined in the main text.
The momentum ($k$) of the maximum of the momentum distribution curve (MDC) of the ARPES
for each band (of index of $\nu$) are noted as $k_{\nu}^{MDC}$
assigned to each frequency of $\omega$ as shown in Eq.\ref{eq:method}.
The quasiparticle spectra of ($\omega$,$k$) is given by
the condition of $det(G^{-1}(k,\omega))=0$
with the condition of Im$\Sigma$=0.
Thus, the variable $|d|$ should be minimized for each frequency.
From the defined reference Hamiltonian of $H(k)$,
the variable $|d|$ is a function of the self-energy for each frequency.
We minimize $|d|$ as a root finding procedure
from the self-energy variable for each frequency.
From this procedure, the self-energy is extracted.

\begin{eqnarray} \label{eq:method}
&d^{2}=\sum_{\nu}|det(G^{-1}(k_{\nu}^{MDC},\omega))|^{2}& \nonumber\\
&\nu=(\alpha, \beta, \gamma, and~\delta)& \nonumber\\
&|d|=f(\Sigma_{xz/yz}(k,\omega),\Sigma_{xy}(k,\omega))&
\end{eqnarray}

Several assumptions
for the simplification of self-energy is made to solve Eq.\ref{eq:method}.
The first, for the low energy, most of orbital character is $xz/yz$ and $xy$.
Thus, we consider the self-energy of $xz/yz$ and $xy$ of Fe($d$) only as shown in Eq.\ref{eq:method}.
The second, we assume that self-energy is $k_{z}$ independent.
The third, for the in-plane momentum dependence of the self-energy,
we employed two ans\"{a}tze, namely, the local $\Sigma$ ansatz and the non-local $\Sigma$ ansatz as shown in Eq.\ref{eq:Self_Energy}

\begin{eqnarray} \label{eq:Self_Energy}
&\Sigma_{m}(k,\omega)&=\Sigma_{m}(K,\omega)~ non-local~\Sigma~ansatz \nonumber\\
&\Sigma_{m}(k,\omega)&=\Sigma_{m}(\omega)~ local~\Sigma~ansatz
\end{eqnarray}

In the local $\Sigma$ ansatz, the self-energy is momentum independent,
and all bands of $\nu$ index for the given frequency
is considered for the minimization of $|d|$ as shown in Eq.\ref{eq:Self_Energy}.
In the non-local $\Sigma$ ansatz, the self-energy is a coarse grained constant in the momentum
space for $K$ of Brillouin zone (BZ) patch as shown in Eq.\ref{eq:Self_Energy}.
We used the 2 dimensional BZ patch in the main text,
in relation to the existing antiferromagnetic (AFM) correlation
with $q$ vector close to $M$.\cite{qureshi2012inelastic}
In practice, the coarse grained momentum $K$=$\Gamma$ includes $k$ for
$\alpha$, $\alpha$' and $\beta$, hole bands, and
the coarse grained momentum $K$=M includes $k$
for $\delta$ and $\gamma$ electron bands.
We consider $\alpha$, $\beta$, $\delta$, and $\gamma$
bands for the extraction of the momentum independent self-energies in the local $\Sigma$ ansatz.
In the non-local $\Sigma$ ansatz, we consider $\alpha$ and $\beta$ bands for $K$=$\Gamma$,
and $\delta$ and $\gamma$ bands for $K$=M, for the extraction of
the $K$ dependent self-energy as described in Eq.\ref{eq:method}.
The description of $\alpha$' could be achieved from
these procedures due to the similar $K$ dependence of self-energy
in between $\alpha$ and $\alpha$' of $xz/yz$ orbital.

From the symmetry of the lattice of LiFeAs, we assumed that self-energy of Fe($d$)
is orbitally diagonal.
From Refs.~\onlinecite{kim2018spin,horvat2017spin,linden2020imaginary}, it is shown that
for the regime of Hund's metal,
due to the larger energy scale of Kondo screening in the orbital sector ($T^{o}_{K}$)
with respect to the SOC ($\lambda_{SOC}$), $T^{o}_{K}>\lambda_{SOC}$,
SOC is not effective on the orbitally diagonal self-energy having a frequency dependent coherence-incoherence crossover.

For orbitally off-diagonal self-energy, in Refs.~\onlinecite{kim2018spin,linden2020imaginary},
it is shown that from the example of Sr$_{2}$RuO$_{4}$, this off-diagonal term
could be absorbed into the renormalization of SOC.
We assume that this orbitally off-diagonal self-energy is zero.
We confirmed that this assumption properly describes the SOC effects on the ARPES near the zero frequency.
Table~\ref{table:SOC} summarizes the effective SOC constant $\lambda_{SOC}$ of the reference Hamiltonian, and
its comparison with the band splitting energy of $\alpha$ and $\alpha'$ band at $k$=$\Gamma$, $\Delta$
for ARPES ($\Delta_{ARPES}$), LDA+SOC ($\Delta_{LDA}$), and LQSGW+SOC ($\Delta_{LQSGW}$).
The $\Delta$ is effectively equal to $Z\lambda_{SOC}$
($Z$ is the renormalization constant of the $xz/yz$ orbital at $k=\Gamma$) where $Z$=1.0 if the self-energy vanishes.

\begin{table*}[h!]
\caption{The SOC induced splitting (in meV unit) of the
$\alpha$ and $\alpha'$ at $k$=$\Gamma$, $\Delta$, of ARPES ($\Delta_{ARPES}$)\cite{borisenko2016direct},
LDA+SOC ($\Delta_{LDA}$), and LQSGW+SOC ($\Delta_{LQSGW}$).
We have shown the effective SOC ($\lambda_{SOC}$) of the reference Hamiltonian, $H(k)$, of the
LDA ($\lambda^{LDA}_{SOC}$) and LQSGW ($\lambda^{LQSGW}_{SOC}$).
The $\Delta$ of the LDA+non-local $\Sigma$ fit ($Z\lambda^{LDA}_{SOC}$) and the LQSGW+non-local $\Sigma$ fit ($Z\lambda^{LQSGW}_{SOC}$)
is also shown, to be compared with its experimental value of $\Delta_{ARPES}$.
}
\begin{tabular}{|| c | c | c | c | c | c | c ||}
\hline
 $\Delta_{ARPES}$\cite{borisenko2016direct} & $\Delta_{LDA}$ & $\Delta_{LQSGW}$ & $\lambda^{LDA}_{SOC}$ & $\lambda^{LQSGW}_{SOC}$ & $Z\lambda^{LDA}_{SOC}$ & $Z\lambda^{LQSGW}_{SOC}$ \\
\hline
 9.5-11.4 & 50 & 35 & 50 & 25 & 12.5 & 9.5 \\
\hline
\end{tabular}
\label{table:SOC}
\end{table*}

\section{V. Determination of $k_{z}$ for the reference of electron sheets for $H^{LDA}$ and $H^{LQSGW}$}

In this section, we describe the
relation in between (i) the variable of $k_{z}$ for electron pockets
for the reference Hamiltonian $H(k)$ and (ii) the spatial locality of self-energy and the validity of the Luttinger's theorem, as discussed in the main text.
The Fermi momentum and the mass of the $xz/yz$ driven electron band ($\delta$ band) computed with both LDA and LQSGW reference Hamiltonians depend strongly on  $k_{z}$ as shown
in Fig.\ref{fig:kz}.   This in turns influences the extracted  self-energy.   Only one value of $k_{z}$ (0.55),
together with the assumption of a $k_{z}$ independent self energy gives a  three dimensional volume consistent with the Luttinger's theorem.

\begin{figure}[ht!]
\includegraphics[width=10cm]{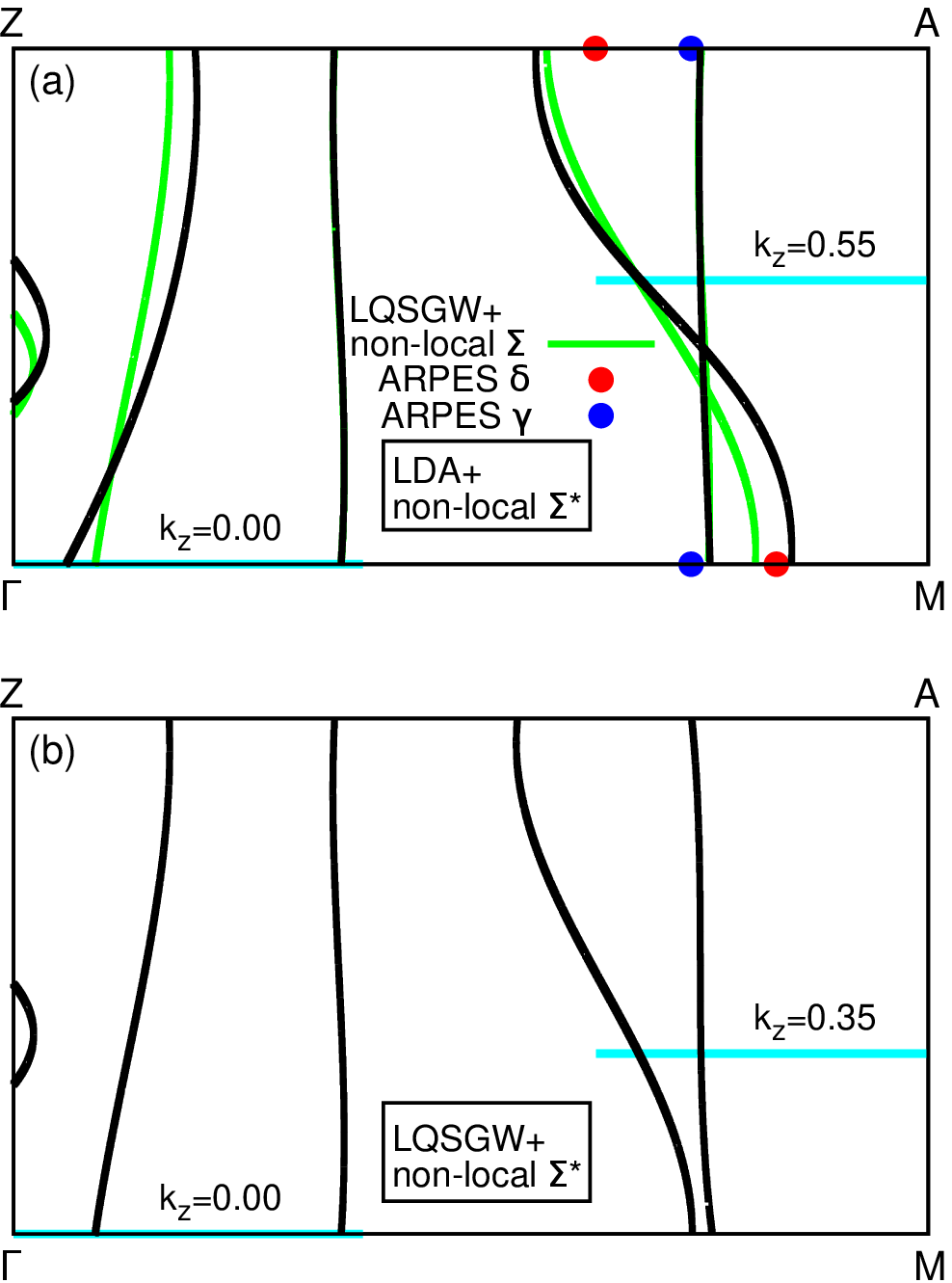}
\caption{(a) Fermi surfaces for the LDA+non-local $\Sigma$ ansatz with $k_{z}$=0.00 for hole bands fit
and $k_{z}$=0.55 for electron bands fit (black color)
(total Fermi surface volume of -0.01 (electrons/unit cell)).
Red and blue dots are the $\delta$ and $\gamma$ Fermi surfaces
measured with ARPES in Ref.\cite{brouet2016arpes}.
Green lines are for the Fermi surface of the LQSGW+non-local $\Sigma$ ansatz with $k_{z}$=0.00 for hole bands fit
and $k_{z}$=0.55 for electron bands fit
(b) Fermi surfaces for the LQSGW+non-local $\Sigma$ ansatz with $k_{z}$=0.00 for hole bands fit
and $k_{z}$=0.35 for electron bands fit
(total Fermi surface volume of +0.18 (electrons/unit cell)).
The ARPES data for fitting is taken from Ref.\cite{miao2016orbital}
(See main text).
\label{fig:FS}
}
\end{figure}

$LDA~reference$ - Table~\ref{table:Linear_LDA_DCA_diff_kz} presents
the self-energy from the LDA+non-local $\Sigma$ ansatz with $k_{z}$=0.00 for hole pockets
and $k_{z}$=0.55 for electron pockets.
It is shown that the self-energy of this case have a
strong spatial non-locality with strong momentum (K) dependence. This feature implies that
the LDA+local $\Sigma$ ansatz is not applicable with $k_{z}$=0.55 for electron pockets.
With this choice of $k_{z}$=0.55 for electron pockets,
the LDA+non-local $\Sigma$ ansatz fulfills the Luttinger's theorem
as shown in Table~\ref{table:FS_volumesS} and Fig.\ref{fig:FS}(a).
In Fig.\ref{fig:FS}(a), it is shown that the LDA+non-local $\Sigma$ ansatz
with $k_{z}$=0.55 for electron pockets provides
an overestimation of total $k_{z}$ dependent dispersion
of the $\delta$ Fermi surface with respect to the ARPES and the LQSGW+non-local $\Sigma$
ansatz with $k_{z}=0.55$ for electron pockets.\cite{brouet2016arpes}
In the main text, we have shown that the setting of $k_{z}$=0.35 for electron
pockets provides more local static self-energy ($\Sigma(0)$) but violates the Luttinger's theorem (see Table~\ref{table:FS_volumesS} and Table~\ref{table:Linear_LDA_DCA_kz_error}).

In summary, for the LDA reference, there is no $k_{z}$ for electron sheet
such that fulfills both the locality of self-energy and the Luttinger's theorem together.

$LQSGW~reference$ - Table~\ref{table:Linear_LQSGW_DCA_diff kz} presents
the self-energy from the LQSGW+non-local $\Sigma$ ansatz with the $k_{z}$=0.00 for hole pockets
and the $k_{z}$=0.35 for electron pockets.
It is shown that the self-energy of this case have
a strong spatial non-locality with strong momentum (K) dependence.
With this choice of the $k_{z}$=0.35 for electron pockets,
the LQSGW+non-local $\Sigma$ ansatz violates the Luttinger's theorem
as shown in Table~\ref{table:FS_volumesS} and Fig.\ref{fig:FS}(b).
In the main text, we have shown that for the setting of $k_{z}$=0.55 for electron pockets,
the Luttinger's theorem is obeyed and the self-energy is spatially local
(see Table~\ref{table:FS_volumesS} and Table~\ref{table:Linear_LQSGW_DCA_kz_error}).

In summary, for the LQSGW reference, there is single value
of $k_{z}$=0.55 for electron pockets
which (i) fulfills the Luttinger's theorem and (ii) provides a validity of the spatially local self-energy.

\begin{table*}[ht!]
\caption{
Zero frequency self-energy ($\Sigma_{m}(K,0)$)
and quasiparticle residue (Z$_{m}$(K)) from the ARPES of LiFeAs
with the LDA reference using the non-local $\Sigma$ ansatz.
We set $k_{z}$=0.00 for K=$\Gamma$ (for hole sheets) and $k_{z}$=0.55 for K=M (for electron sheets), for the LDA reference.
Error bars (total) are computed from the peak width of both in plane $k$ and
out of plane $k_{z}$.
For the peak width, we adapted the resolution in the MDC of LiFeAs, 0.01 ($\pi/a$) (0.008 ${\AA}^{-1}$).
Here, $\sqrt{2}a$ is the $a_{lattice}$, where $a_{lattice}$ is the lattice constant of LiFeAs.
This resolution in the MDC is taken from the best resolution
limit for the $\beta$ sheet with the resolution limit of the energy of 3 meV.
}
\begin{tabular}{|| c || c | c | c | c ||}
\hline
 ~ & $\Sigma_{m}(\Gamma,0)$ (eV)  & Z$_{m}$($\Gamma$) & $\Sigma_{m}(M,0)$ (eV) & Z$_{m}$(M) \\
\hline
$xy$ & 0.029$\pm$0.025 & 0.15$\pm$0.01 & -0.153$\pm$0.060 & 0.10$\pm$0.01 \\
\hline
$xz/yz$ & -0.083$\pm$0.040 & 0.25$\pm$0.13 & 0.042$\pm$0.014 & 0.34$\pm$0.04 \\
\hline
\end{tabular}
\label{table:Linear_LDA_DCA_diff_kz}
\end{table*}

\begin{table*}[ht!]
\caption{The Fermi surface volumes (electrons/unit cell)
in (a) the LDA, (b) the non-local $\Sigma$ fitting with the reference of the LDA with $k_{z}$=0.35 for electron bands,
(c) the local $\Sigma$ fitting with the reference of the LDA with $k_{z}$=0.35 for electron bands,
(d) the non-local $\Sigma$ fitting with the reference of the LDA with $k_{z}$=0.55 for electron bands (See Fig.\ref{fig:FS}(a)),
(e) the LQSGW,
(f) the non-local $\Sigma$ fitting with the reference of LQSGW with $k_{z}$=0.55 for electron bands, and
(g) the results of the LQSGW+non-local $\Sigma$ fit with $k_{z}$=0.35 for electron bands
which violates the Luttinger's theorem (See Fig.\ref{fig:FS}(b)).
We fit the hole bands with $k_{z}$=0.00 for both the LDA and the LQSGW references.
}
\begin{tabular}{|| c || c | c | c | c | c | c ||}
\hline
 ~ & $\alpha$' & $\alpha$ & $\beta$ & $\gamma$ & $\delta$ & total \\
\hline
 LDA & 0.01 & 0.14 & 0.33 & 0.18 & 0.28 & -0.02 \\
\hline
 LDA+non-local $\Sigma$ ($k_{z}$=0.35 for fitting of electron bands)& 0.00 & 0.08 & 0.37 & 0.22 & 0.39 & +0.17 \\
\hline
 LDA+local $\Sigma$ ($k_{z}$=0.35 for fitting of electron bands) & 0.00 & 0.06 & 0.36 & 0.19 & 0.35 & +0.12 \\
\hline
 LDA+non-local $\Sigma$ ($k_{z}$=0.55 for fitting of electron bands)& 0.00 & 0.08 & 0.37 & 0.19 & 0.24 & -0.01 \\
\hline
 LQSGW & 0.00 & 0.08 & 0.35 & 0.20 & 0.21 & -0.03 \\
\hline
 LQSGW+non-local $\Sigma$ ($k_{z}$=0.55 for fitting of electron bands) & 0.00 & 0.05 & 0.36 & 0.20 & 0.26 & +0.04 \\
\hline
 LQSGW+non-local $\Sigma$ ($k_{z}$=0.35 for fitting of electron bands) & 0.00 & 0.05 & 0.36 & 0.23 & 0.37 & +0.18 \\
\hline
\end{tabular}
\label{table:FS_volumesS}
\end{table*}

\begin{table*}[ht!]
\caption{
Zero frequency self-energy ($\Sigma_{m}(K,0)$)
and quasiparticle residue (Z$_{m}$(K)) from the ARPES of LiFeAs
with the LQSGW reference using the non-local $\Sigma$ ansatz.
We set $k_{z}$=0.00 for K=$\Gamma$ (for hole sheets) and $k_{z}$=0.35 for K=M (for electron sheets), for the LQSGW reference.
Error bars (total) are computed from the peak width of both in plane $k$ and
out of plane $k_{z}$.
For the peak width, we adapted the resolution in the MDC of LiFeAs, 0.01 ($\pi/a$) (0.008 ${\AA}^{-1}$).
Here, $\sqrt{2}a$ is the $a_{lattice}$, where $a_{lattice}$ is the lattice constant of LiFeAs.
This resolution in the MDC is taken from the best resolution
limit for the $\beta$ sheet with the resolution limit of the energy of 3 meV.
}
\begin{tabular}{|| c || c | c | c | c ||}
\hline
 ~ & $\Sigma_{m}(\Gamma,0)$ (eV)  & Z$_{m}$($\Gamma$) & $\Sigma_{m}(M,0)$ (eV) & Z$_{m}$(M) \\
\hline
$xy$ & 0.002$\pm$0.014 & 0.21$\pm$0.01 & 0.041$\pm$0.033 & 0.21$\pm$0.01 \\
\hline
$xz/yz$ & -0.027$\pm$0.003 & 0.38$\pm$0.01 & -0.188$\pm$0.021 & 0.16$\pm$0.01 \\
\hline
\end{tabular}
\label{table:Linear_LQSGW_DCA_diff kz}
\end{table*}

\begin{table*}[ht!]
\caption{
Zero frequency self-energy ($\Sigma_{m}(K,0)$)
and quasiparticle residue (Z$_{m}$(K)) from the ARPES of LiFeAs
with the LDA reference using the non-local $\Sigma$ ansatz.
We set $k_{z}$=0.00 for K=$\Gamma$ (for hole sheets) and $k_{z}$=0.35 for K=M (for electron sheets), for the LDA reference.
Error bars are computed from the peak width of in-plane $k$ (parentheses is out of plane $k_{z}$).
For the peak width, we adapted the resolution in the MDC of LiFeAs, 0.01 ($\pi/a$) (0.008 ${\AA}^{-1}$).
Here, $\sqrt{2}a$ is the $a_{lattice}$, where $a_{lattice}$ is the lattice constant of LiFeAs.
This resolution in the MDC is taken from the best resolution
limit for the $\beta$ sheet with the resolution limit of the energy of 3 meV.
}
\begin{tabular}{|| c || c | c | c | c ||}
\hline
 ~ & $\Sigma_{m}(\Gamma,0)$ (eV)  & Z$_{m}$($\Gamma$) & $\Sigma_{m}(M,0)$ (eV) & Z$_{m}$(M) \\
\hline
$xy$ & 0.029$\pm$0.025(0.000) & 0.15$\pm$0.01(0.00) & -0.130$\pm$0.062(0.004) & 0.12$\pm$0.01(0.01) \\
\hline
$xz/yz$ & -0.083$\pm$0.040(0.000) & 0.25$\pm$0.13(0.01) & -0.113$\pm$0.013(0.023) & 0.16$\pm$0.01(0.02) \\
\hline
\end{tabular}
\label{table:Linear_LDA_DCA_kz_error}
\end{table*}

\begin{table*}[ht!]
\caption{
Zero frequency self-energy ($\Sigma_{m}(K,0)$)
and quasiparticle residue (Z$_{m}$(K)) from the ARPES of LiFeAs
with the LQSGW reference using the non-local $\Sigma$ ansatz.
We set $k_{z}$=0.00 for K=$\Gamma$ (for hole sheets) and $k_{z}$=0.55 for K=M (for electron sheets), for the LQSGW reference.
Error bars are computed from the peak width of in-plane $k$ (parentheses is out of plane $k_{z}$).
For the peak width, we adapted the resolution in the MDC of LiFeAs, 0.01 ($\pi/a$) (0.008 ${\AA}^{-1}$).
Here, $\sqrt{2}a$ is the $a_{lattice}$, where $a_{lattice}$ is the lattice constant of LiFeAs.
This resolution in the MDC is taken from the best resolution
limit for the $\beta$ sheet with the resolution limit of the energy of 3 meV.
}
\begin{tabular}{|| c || c | c | c | c ||}
\hline
 ~ & $\Sigma_{m}(\Gamma,0)$ (eV)  & Z$_{m}$($\Gamma$) & $\Sigma_{m}(M,0)$ (eV) & Z$_{m}$(M) \\
\hline
$xy$ & 0.002$\pm$0.014(0.000) & 0.21$\pm$0.01(0.00) & 0.044$\pm$0.036(0.001) & 0.18$\pm$0.00(0.01) \\
\hline
$xz/yz$ & -0.027$\pm$0.003(0.000) & 0.38$\pm$0.01(0.00) & -0.051$\pm$0.006(0.114) & 0.30$\pm$0.03(0.03) \\
\hline
\end{tabular}
\label{table:Linear_LQSGW_DCA_kz_error}
\end{table*}

\section{VI. Error bar in the self-energy, in plane $k$ error versus out of plane $k_{z}$ error}

In this section, we describe the procedure for the extraction of the error bar
in the self-energy.

In the extraction of error bar of
zero frequency self-energy ($\Sigma_{m}(K,0)$)
and quasiparticle residue (Z$_{m}$(K)) from the ARPES,
we consider two independent sources of error
(i) in plane $k$ error and (ii) out of plane $k_{z}$ error,
with consideration of the MDC peak width of 0.01 ($\pi/a$) (0.008 ${\AA}^{-1}$).
There is an intrinsic uncertainty in the value of $k_{z}$
as the surface breaks translation symmetry and
therefore, the $k_{z}$ is not a good quantum number.
As shown in Fig.\ref{fig:kz}, the most dominant $k_{z}$ dependent
variation is in the $xz/yz$ orbital dominant $\delta$ band placed at $K$=M,
and other bands have a smaller $k_{z}$ dependency.
As a results, the most of the error bars are from the
in plane $k$ error, and a sizable $k_{z}$ peak width
induced error is only at the $K$=M for $xz/yz$ orbital (contribute to the $\delta$ band),
for both the LDA reference and the LQSGW reference,
as shown in Table~\ref{table:Linear_LDA_DCA_kz_error} and Table~\ref{table:Linear_LQSGW_DCA_kz_error}.
The large Fermi velocity of the $\gamma$ band results in
the large error bar in the zero frequency self-energy of the $xy$ orbital at $K$=M from the in plane $k$ error.

\section{VII. Locality of dynamical self-energy}

In this section, we present dynamical part self-energies
in the LDA reference and LQSGW reference, in the main text,
with $k_{z}$=0.35 for electron bands fit in the LDA reference
and $k_{z}$=0.55 for electron bands fit in the LQSGW reference.

\begin{figure}[ht!]
\includegraphics[width=12cm]{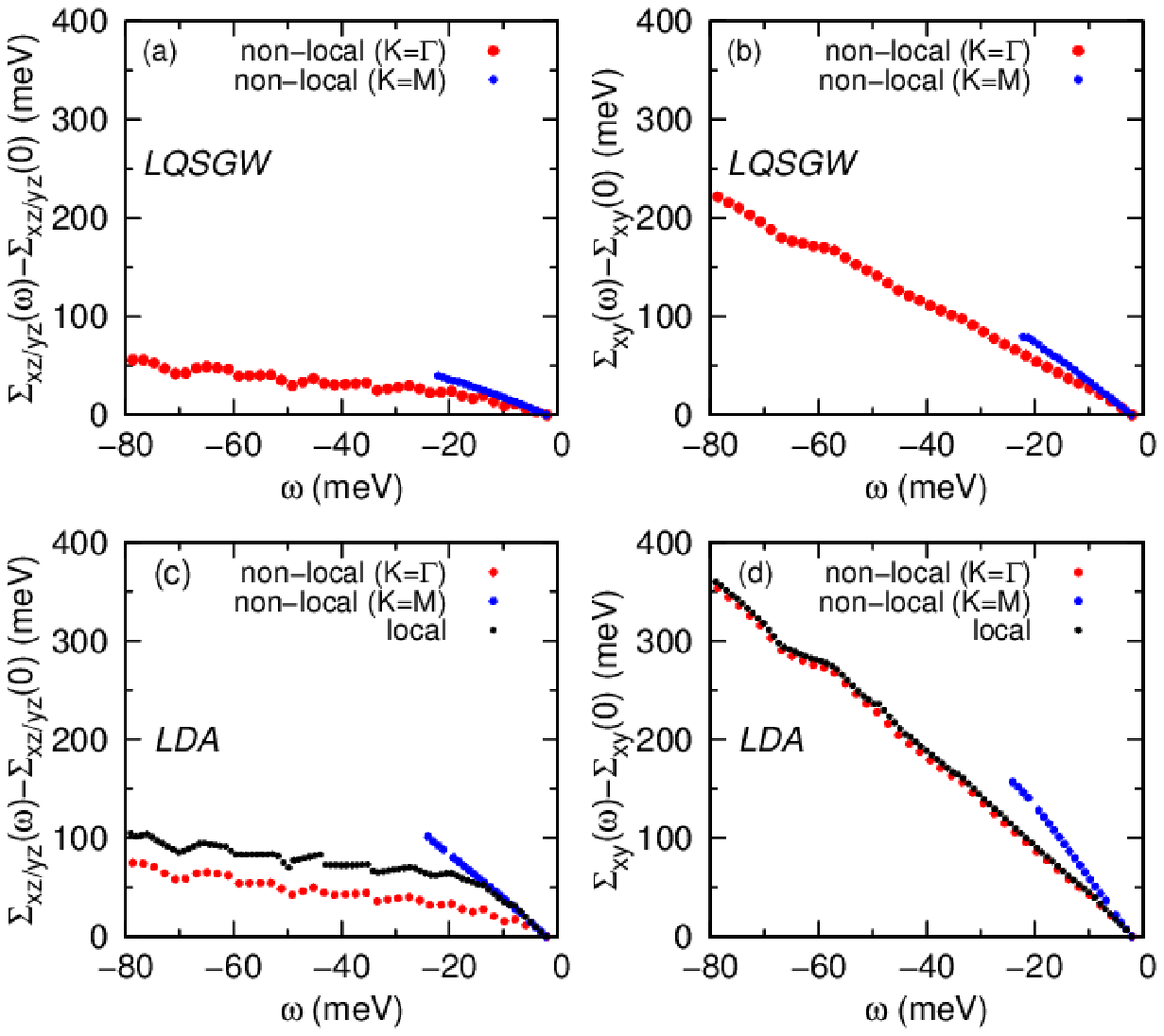}
\caption{(a) and (b)
Dynamical part of the self energies,
$\Sigma_{m}(K,\omega)-\Sigma_{m}(K,0)$
in the LQSGW+non-local $\Sigma$ ansatz for $xz/yz$ orbital and $xy$ orbital, respectively.
(c) and (d) Dynamical part self energy of $xz/yz$ and $xy$ orbitals,
in the LDA+non-local $\Sigma$ ansatz and in the LDA+local $\Sigma$ ansatz.
Here, we use $k_{z}$=0.00 for hole bands
for both the LDA and the LQSGW references.
We use $k_{z}$=0.35 for electron bands
for the LDA references.
We use $k_{z}$=0.55 for electron bands
for the LQSGW references (see the main text and the Section V).
\label{fig:SE}
}
\end{figure}

The full frequency dependence of the extracted self-energies is presented on Fig.~\ref{fig:SE}, which displays the
dynamical part $\mathrm{Re}\Sigma_m(K,\omega)-\mathrm{Re}\Sigma_m(K,0)$ for the different schemes considered in this work.
By comparing the $K=\Gamma$ and $K=M$ data, it is immediately apparent from this figure that the fits based
on the LQSGW reference lead to a much higher degree of locality for both orbital components
than those based on the LDA reference.
These factors imply that the self-energy from the LQSGW referencce is spatially local,
confirms the validity of the LQSGW+local $\Sigma$ ansatz for the description of
the quasiparticle of LiFeAs.

\section{VIII. Validity of the LQSGW+non-local $\Sigma$ ansatz for ARPES}

\begin{figure}[ht!]
\includegraphics[width=16.5cm]{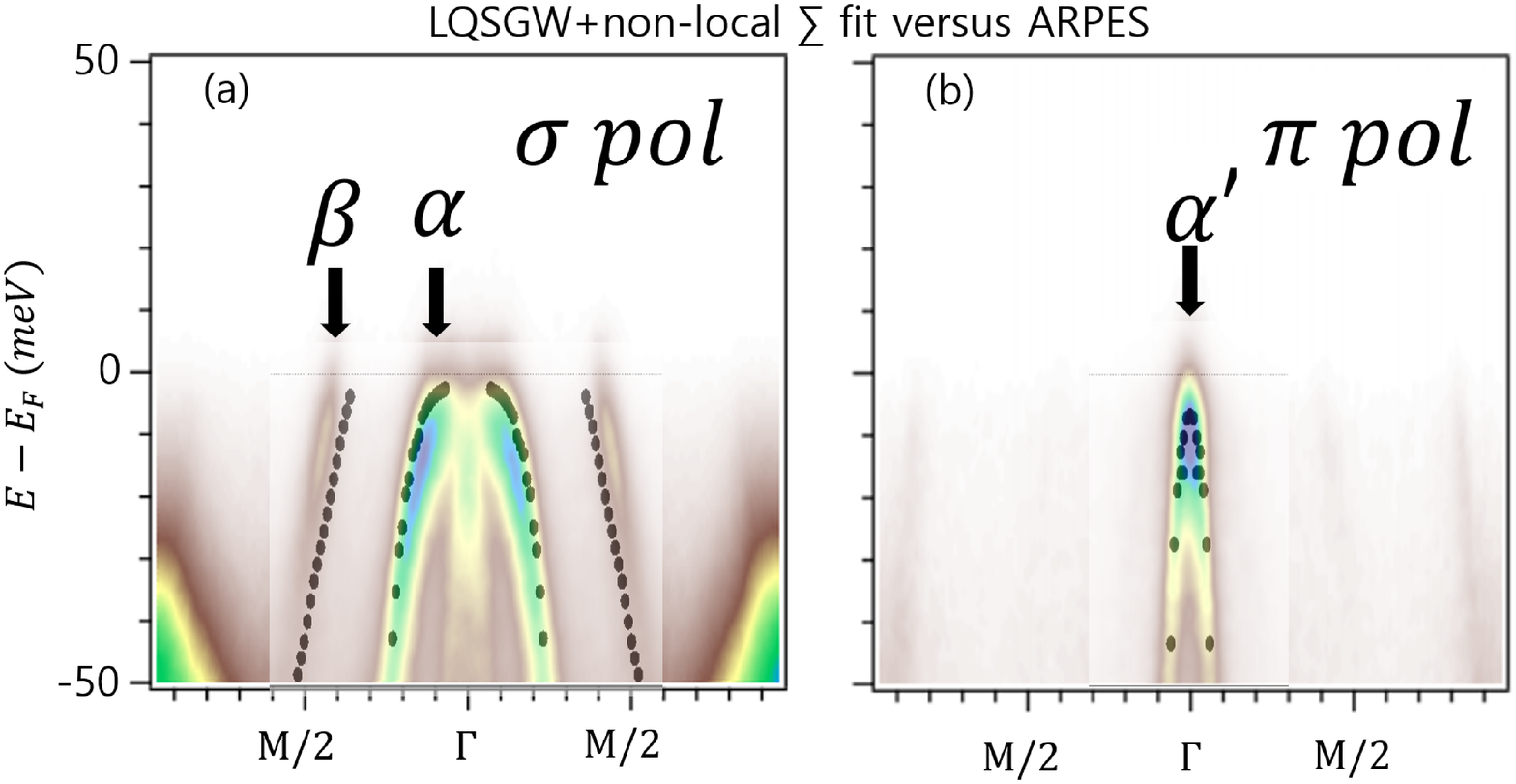}
\caption{(a) Comparison of band dispersions of LiFeAs
of Ref.\cite{miao2016orbital}
from the ARPES (color contour) and the LQSGW+non-local $\Sigma$ ansatz
for hole pockets ($\alpha$ and $\beta$) in the $\Gamma$-M $k$ path ($k_{z}$=0.00)
from $\sigma$ polarized light.
(b) Same as (a) for $\alpha'$ from $\pi$ polarized light.\cite{miao2016orbital}
\label{fig:ARPES1}
}
\end{figure}

\begin{figure}[ht!]
\includegraphics[width=16.5cm]{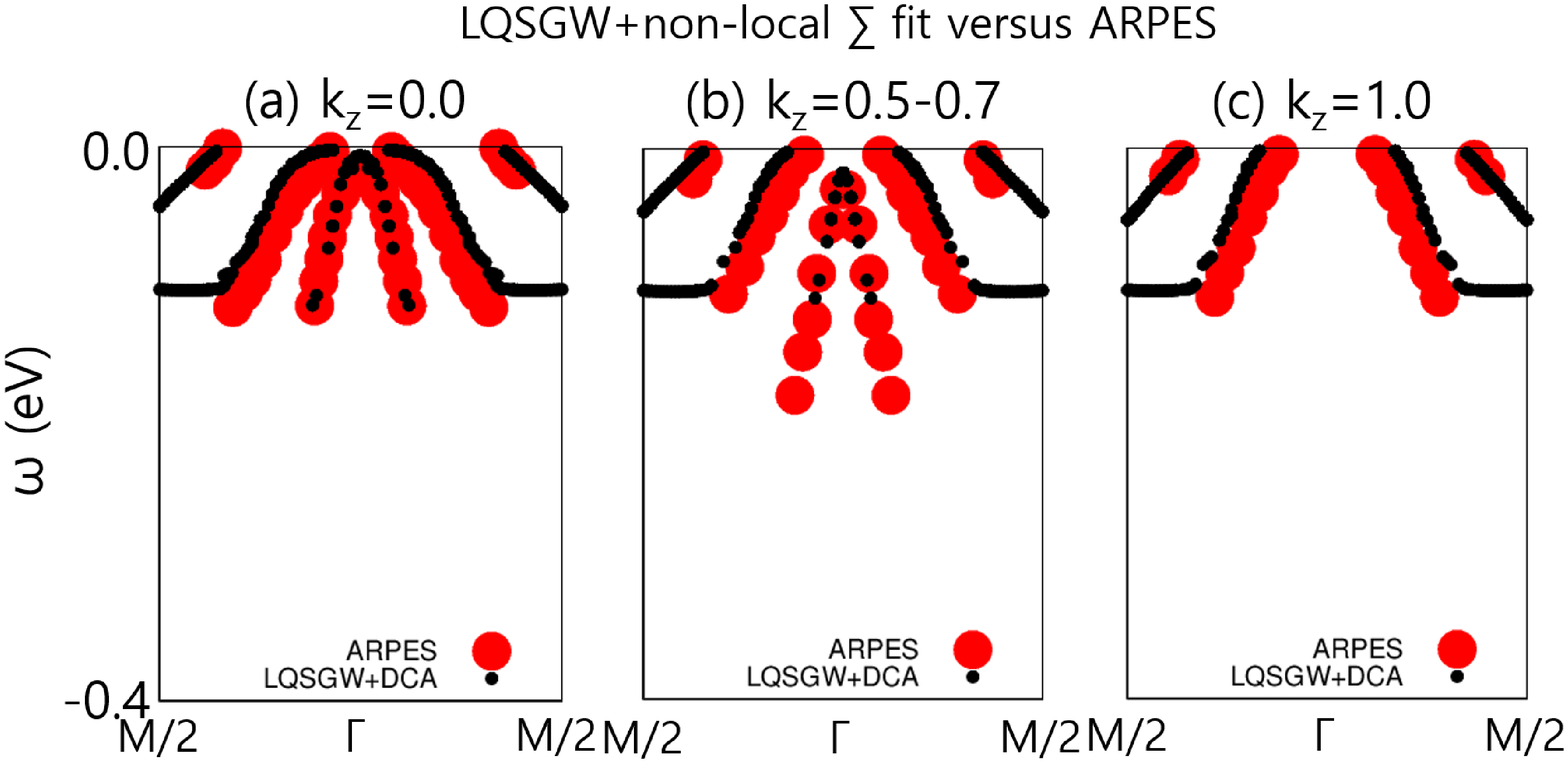}
\caption{(a) Comparison of band dispersions of LiFeAs
of Ref.\cite{wang2015topological} from the ARPES (red circle)
and the LQSGW+non-local $\Sigma$ ansatz (black dots) for $k_{z}$=0.00
for hole pockets ($\alpha$, $\alpha'$, and $\beta$) in the $\Gamma$-M $k$ path ($k_{z}$=0.00).
(b) Same as (a) for $k_{z}$=0.50-0.70\cite{wang2015topological}.
(c) Same as (a) for $k_{z}$=1.00\cite{wang2015topological}.
\label{fig:ARPES2}
}
\end{figure}

To show the validity of the extracted self-energy
of the LQSGW+non-local $\Sigma$ ansatz from the quasiparticle
of the ARPES experiment in the main text,
we compared spectra from the LQSGW+non-local $\Sigma$ ansatz to the ARPES of Ref.\cite{miao2016orbital}
in Fig.\ref{fig:ARPES1}
and the ARPES of Ref.\cite{wang2015topological}
in Fig.\ref{fig:ARPES2}
with different $k$ path ($\Gamma$-M) and $k_{z}$ values.
Fig.\ref{fig:ARPES1} and Fig.\ref{fig:ARPES2}
implies that present LQSGW+non-local $\Sigma$ ansatz
provides a good description of the quasiparticle
spectra for multiple examples of the $k$ path and the $k_{z}$.

\section{IX. Spin orbit coupling of LiFeAs}

\begin{figure}[ht!]
\includegraphics[width=16cm]{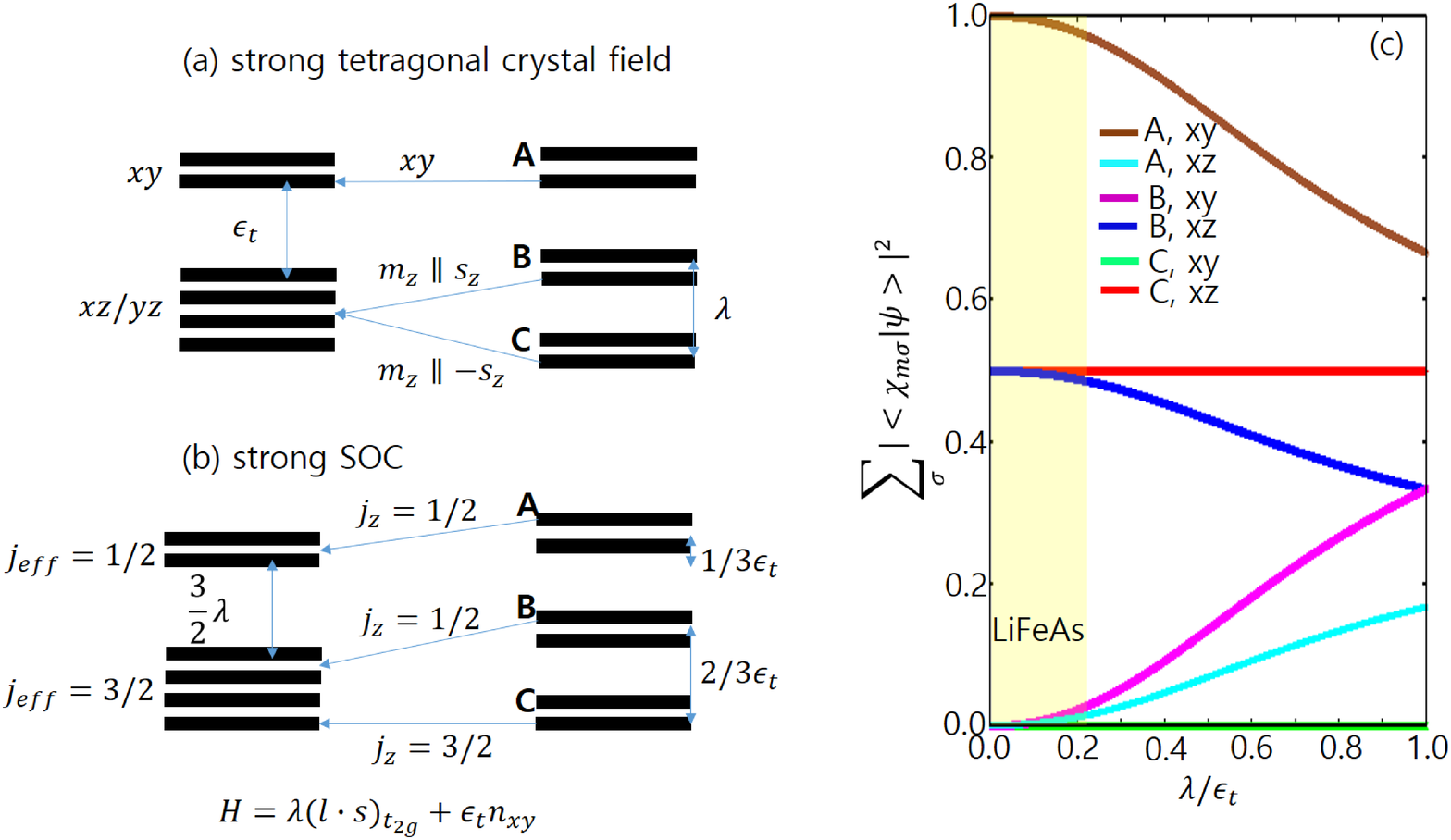}
\caption{(a) Multiplet states in the presence of strong
tetragonal crystal field ($\varepsilon_{t}$),
with small spin orbit coupling ($\lambda$).
(b) Multiplet states in the presence of strong
spin orbit coupling ($\lambda$)
small tetragonal crystal field ($\varepsilon_{t}$).
A, B, and C Kramers doublets are continuously
connected from the regime of (a) to the regime of (b),
presented in Eq.\ref{eq:large_CF} and Eq.\ref{eq:large_SOC}.
(c) Orbital contents of A, B, and C Kramers doublet
with variation of $\lambda/\varepsilon_{t}$.
$\psi$ is the Kramer's doublet A, B, and C
of Eq.\ref{eq:large_CF} and Eq.\ref{eq:large_SOC}.
$\chi_{m,\sigma}$ is the atomic $t_{2g}$ orbital
($m$ : orbital, $\sigma$ : spin).
From the tetragonal symmetry, the $yz$ orbital
have a same projection weight with respect to the $xz$ orbital.
The regime for non-degenerate bands of LiFeAs
is denoted from yellow color.
\label{fig:SOC_CF}
}
\end{figure}

In this section  we  estimate  the
admixture of
 $xz/yz$ and $xy$   orbital character near the Fermi level
 near  the $\Gamma $ point.


Fig.\ref{fig:SOC_CF} presents single particle
$t_{2g}$ states in the presence of a tetragonal crystal field
($\varepsilon_{t}$) and SOC ($\lambda$)
as defined  by  the one site  Hamiltonian in Eq.\ref{eq:Hamiltonian}.

\begin{eqnarray} \label{eq:Hamiltonian}
&H=&\lambda(l \cdot s)_{t_{2g}} + \frac{1}{2}\varepsilon_{t}(c_{xy,\uparrow}^{\dag}c_{xy,\uparrow}+c_{xy,\downarrow}^{\dag}c_{xy,\downarrow}-c_{xz,\uparrow}^{\dag}c_{xz,\uparrow}-c_{xz,\downarrow}^{\dag}c_{xz,\downarrow}-c_{yz,\uparrow}^{\dag}c_{yz,\uparrow}-c_{yz,\downarrow}^{\dag}c_{yz,\downarrow})
\end{eqnarray}

The $\lambda(l \cdot s)_{t_{2g}}$ of Eq.\ref{eq:Hamiltonian} is given by Eq.\ref{eq:Hamiltonian_SOC},
in the order of $xz,\uparrow$, $yz,\uparrow$, $xy,\uparrow$, $xz,\downarrow$, $yz,\downarrow$, and $xy,\downarrow$ states.

\begin{eqnarray} \label{eq:Hamiltonian_SOC}
\lambda(l \cdot s)_{t_{2g}}=\begin{pmatrix}
0 & -i\frac{\lambda}{2} & 0 & 0 & 0 & i\frac{\lambda}{2} \\
i\frac{\lambda}{2} & 0 & 0 & 0 & 0 & -\frac{\lambda}{2} \\
0 & 0 & 0 & -i\frac{\lambda}{2} & \frac{\lambda}{2} & 0 \\
0 & 0 & i\frac{\lambda}{2} & 0 & i\frac{\lambda}{2} & 0 \\
0 & 0 & \frac{\lambda}{2} & -i\frac{\lambda}{2} & 0 & 0 \\
-i\frac{\lambda}{2} & -\frac{\lambda}{2} & 0 & 0 & 0 & 0
\end{pmatrix}
\end{eqnarray}




The strong and weak  $\lambda$ cases
are shown  in Fig.\ref{fig:SOC_CF}(b)  and Fig.\ref{fig:SOC_CF}(a), respectively.
The eigenvectors of the Hamiltonian  Eq.\ref{eq:Hamiltonian},
are Kramer's doublets A, B, and C.   Their orbital content is shown schematically   as a function of $\lambda \over \varepsilon_{t} $ in Fig.\ref{fig:SOC_CF}.

\begin{eqnarray} \label{eq:large_CF}
&|m_{z}||s_{z},s_{z}=\uparrow>=&-\frac{1}{\sqrt{2}}(|xz,\uparrow>+i|yz,\uparrow>) \nonumber\\
&|m_{z}||s_{z},s_{z}\downarrow>=&\frac{1}{\sqrt{2}}(|xz,\downarrow>-i|yz,\downarrow) \nonumber\\
&|m_{z}||-s_{z},s_{z}=\downarrow>=&-\frac{1}{\sqrt{2}}(|xz,\downarrow>+i|yz,\downarrow>) \nonumber\\
&|m_{z}||-s_{z},s_{z}=\uparrow>=&\frac{1}{\sqrt{2}}(|xz,\uparrow>-i|yz,\uparrow) \nonumber\\
&|xy,\uparrow>& \nonumber\\
&|xy,\downarrow>&
\end{eqnarray}

\begin{eqnarray} \label{eq:large_SOC}
|j_{eff}=1/2,j_{z}=1/2> \nonumber\\
=-\frac{1}{\sqrt{3}}(|xy,\uparrow>+|yz,\downarrow>+i|xz,\downarrow>) \nonumber\\
|j_{eff}=1/2,j_{z}=-1/2> \nonumber\\
=\frac{1}{\sqrt{3}}(|xy,\downarrow>-|yz,\uparrow>+i|xz,\uparrow>) \nonumber\\
|j_{eff}=3/2,j_{z}=1/2> \nonumber\\
=\frac{1}{\sqrt{6}}(2|xy,\uparrow>-|yz,\downarrow>-i|xz,\downarrow>) \nonumber\\
|j_{eff}=3/2,j_{z}=-1/2> \nonumber\\
=\frac{1}{\sqrt{6}}(2|xy,\downarrow>+|yz,\uparrow>-i|xz,\uparrow>) \nonumber\\
|j_{eff}=3/2,j_{z}=3/2> \nonumber\\
=-\frac{1}{\sqrt{2}}(|yz,\uparrow>+i|xz,\uparrow>) \nonumber\\
|j_{eff}=3/2,j_{z}=-3/2> \nonumber\\
=\frac{1}{\sqrt{2}}(|yz,\downarrow>-i|xz,\downarrow>)
\end{eqnarray}

The single particle states  A, B, and C of Eq.\ref{eq:large_CF} and
Eq.\ref{eq:large_SOC} is connected with the variable
of $\lambda/\varepsilon_{t}$.
Fig.\ref{fig:SOC_CF}(c) presents orbital contents
in the $t_{2g}$ orbital in the presence of the tetragonal crystal field
($\varepsilon_{t}$) and SOC ($\lambda$),
as a function of $\lambda/\varepsilon_{t}$.

In the region of parameters that correspond to  the $\Gamma$ point of  LiFeAs,
$\varepsilon_{t}$ correspond to  $\sim$250 meV $\lambda_{SOC}$=50 meV, within LDA,
hence the ratio $\lambda_{SOC}/\varepsilon_{t}$ is less than
0.2 for LiFeAs.This estimate of the upper limit of $\lambda_{SOC}/\varepsilon_{t}$
also holds for LQSGW+$\lambda_{SOC}$.

\section{X. Transport properties of LiFeAs and the LQSGW+local $\Sigma$ ansatz}

In this section, we analyze transport data to show that a
spatially local scattering rate of LiFeAs, motivated by the  LQSGW+local $\Sigma$ ansatz,
provides a natural explanation  for the  transport experiments of Ref.\cite{rullier2012multiorbital}.
We can view this as an independent corroboration, that  the  scattering rate is
spatially local,
while strongly orbital dependent can be taken to be the same in the electron and hole pockets.

\begin{figure}[h!]
\includegraphics[width=10cm]{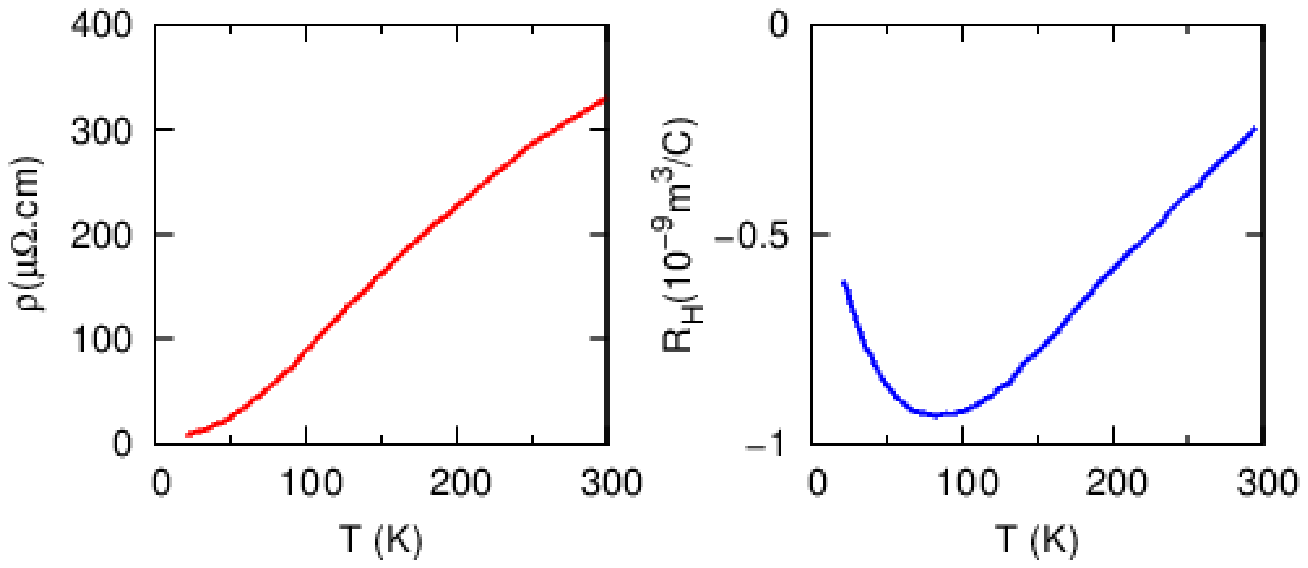}
\caption{Experimental temperature dependent resistivity ($\rho$) of LiFeAs (left panel),
and experimental temperature dependent Hall coefficient (R$_{H}$) of LiFeAs (right panel),
adapted from Ref.\cite{rullier2012multiorbital}.
\label{fig:sigma}
}
\end{figure}

Fig.\ref{fig:sigma} presents the experimental temperature dependent
resistivity ($\rho$) and the experimental temperature dependent Hall coefficient
adapted from Ref.\cite{rullier2012multiorbital}.
This data shows that the Hall coefficient is electron like,
most prominently around the temperature of $\sim$100 K.
With lowering of the temperature below $\sim$100 K, this electron like
Hall conductivity is reduced.
Around the temperature of $\sim$100 K, the resistivity data shows that
upon cooling from this temperature, a Fermi liquid behaviour
is obtained as $\rho$ proportional to $T^{2}$.

We  use a very simplified model of the transport summarized in
Eq.\ref{eq:transport} expressing the
resistivity and the Hall coefficient in terms of the
mobility $\mu$,  a  charge carrier number $n$
and a  scattering rate (1/$\tau$)  with the mobility  $\mu$, of each FS pocket
given by  Eq.\ref{eq:mobility}.

Assuming  that the  scattering rate (1/$\tau$)  depends on the orbital but
not on whether the pocket is electron  or hole like,    we apply  the scattering rate, (1/$\tau$),
of the $xz/yz$ to $\alpha$ and $\delta$ Fermi surfaces,
while  the scattering rate of the $xy$ is applied to $\beta$ and $\gamma$ Fermi surfaces.
$m_{\nu}$ is the bare mass of each FS which we estimate from our  LQSGW reference as shown in Table~\ref{table:mass}.
$m_{\alpha}$ and $m_{\beta}$
are estimated from  the $\Gamma$-X $k$-path ($k_{z}$=0.00),
while  $m_{\gamma}$ is
estimated from $\Gamma$-M $k$-path ($k_{z}$= 0.55).
While  $m_{\delta}$
has substantial $k_{z}$ dependence as shown in Fig.\ref{fig:kz},  transport quantities involve a weighted average of occupation
and mobility.  In our simplified model, we  use an  average value
1.20$m_{e}$  as  summarized in Table~\ref{table:mass}.

We checked  that  variations in  $m_{\delta}$ in the range of
[1.00$m_{e}$,1.80$m_{e}$] gives rise to
similar trends for the extracted scattering rate,
and the converted transport data.

For the carrier number of each FS, we take the value
from the LQSGW+non-local $\Sigma$ ansatz as shown in Table~\ref{table:mass}.

\begin{table*}[t!]
\caption{Mass of the bare band structure ($m_{\nu}$, $\nu$=$\alpha$, $\beta$, $\gamma$, and $\delta$)
of LQSGW+$\lambda_{SOC}$ in the unit of the free electron mass ($m_{e}$),
and the charge carrier number ($n_{\nu}$) of the LQSGW+non-local $\Sigma$ ansatz
from Table~\ref{table:FS_volumesS} ($k_{z}$=0.55 for electron bands fit).
}
\begin{tabular}{|| c | c | c | c ||}
\hline\hline
$m_{\alpha}$  & $m_{\beta}$ & $m_{\gamma}$ & $m_{\delta}$ \\
\hline
1.73$m_{e}$ & 1.36$m_{e}$ & 0.79$m_{e}$ & 1.20$m_{e}$ \\
\hline\hline
$n_{\alpha}$  & $n_{\beta}$ & $n_{\gamma}$ & $n_{\delta}$ \\
\hline
0.05 & 0.36 & 0.20 & 0.26\\
\hline\hline
\end{tabular}
\label{table:mass}
\end{table*}

\begin{eqnarray} \label{eq:transport}
&\rho^{-1}(T)=&\sum_{\nu}n_{\nu}e\mu_{\nu} \nonumber\\
&R_{H}(T)=&(\sum_{\nu}n_{\nu}e\mu_{\nu}^{2}\times sgn(\nu))/\sigma^{2},~sgn(\nu)\textrm{=+1~for~holes,~-1~for~electrons} \nonumber\\
\end{eqnarray}

\begin{eqnarray} \label{eq:mobility}
&\mu_{\alpha}=&\frac{e\tau_{xz/yz}}{m_{\alpha}} \nonumber\\
&\mu_{\beta}=&\frac{e\tau_{xy}}{m_{\beta}} \nonumber\\
&\mu_{\gamma}=&\frac{e\tau_{xy}}{m_{\gamma}} \nonumber\\
&\mu_{\delta}=&\frac{e\tau_{xz/yz}}{m_{\delta}}
\end{eqnarray}

\begin{figure}[t!]
\includegraphics[width=9cm]{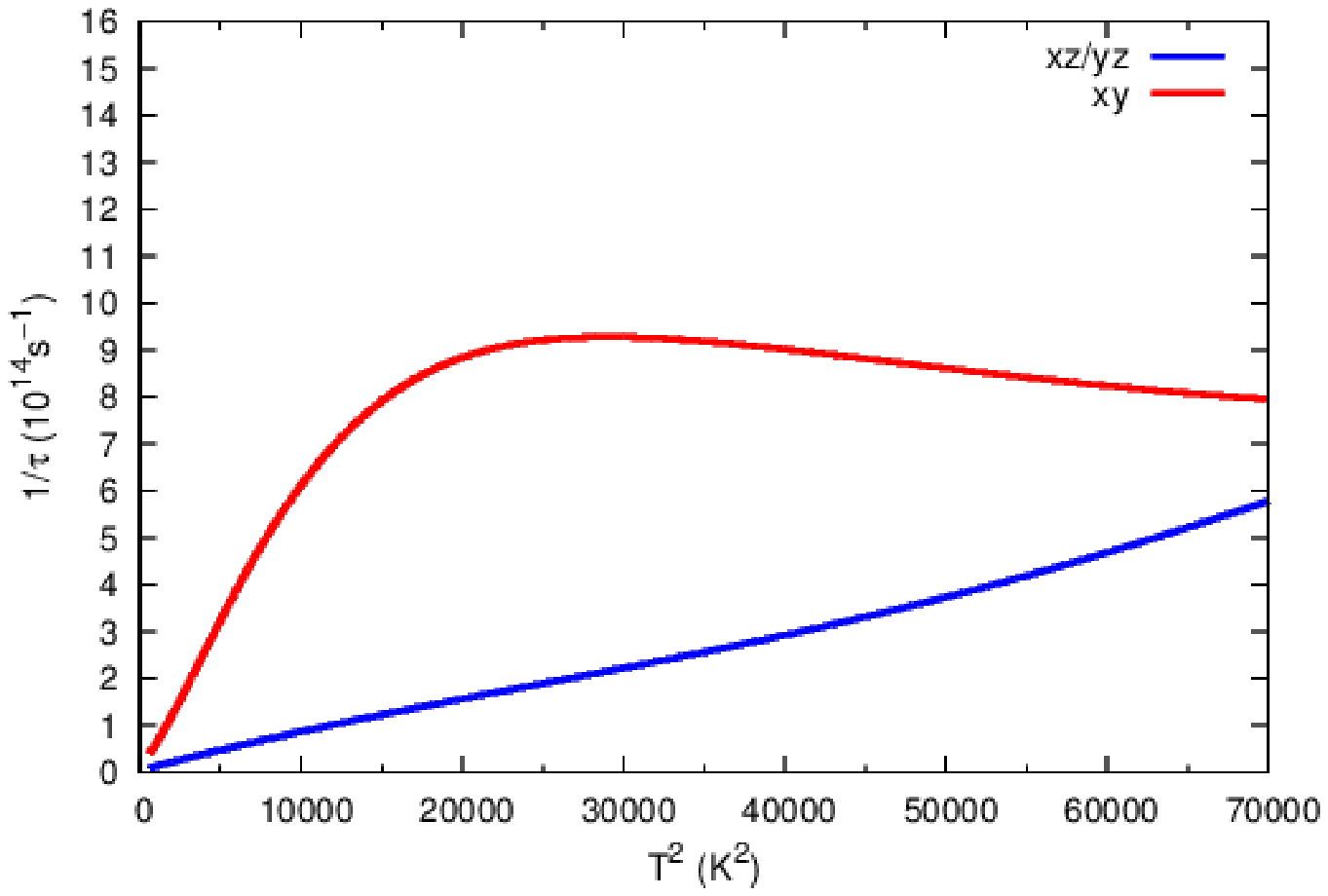}
\caption{Extracted scattering rate ($1/\tau$) of $xz/yz$ and $xy$ orbitals
from the data of Fig.\ref{fig:sigma},\cite{rullier2012multiorbital}
and using Eq.\ref{eq:transport}, Eq.\ref{eq:mobility}, and Table~\ref{table:mass}.
\label{fig:tau}
}
\end{figure}

\begin{figure}[t!]
\includegraphics[width=9cm]{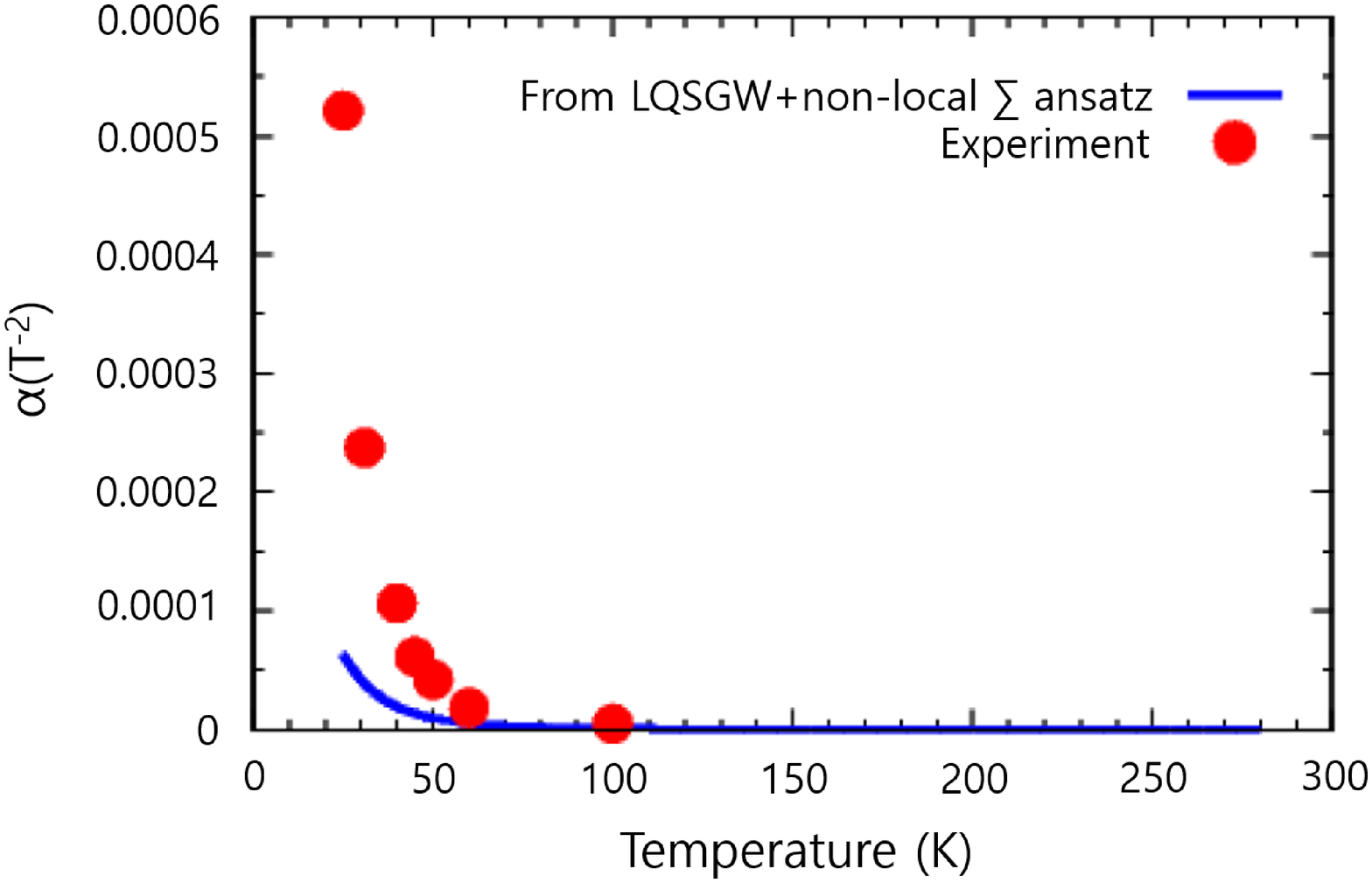}
\caption{Computed temperature dependent
magnetoresistance coefficient ($\alpha$)
from Eq.\ref{eq:alpha} and Eq.\ref{eq:mobility}, using the extracted
scattering rate $1/\tau$ of Fig.\ref{fig:tau}
with Table~\ref{table:mass}
of the LQSGW+non-local $\Sigma$ ansatz (blue line).
The computed $\alpha$ is compared with its experimental
value adapted from Ref.\cite{rullier2012multiorbital} (red dots).
\label{fig:alpha}
}
\end{figure}

Fig.\ref{fig:tau} presents extracted
temperature dependent scattering rate ($1/\tau$)
of the $xz/yz$ and the $xy$ orbitals from the experimental data of
$\rho$ and R$_{H}$ of Ref.\cite{rullier2012multiorbital}.
It is shown that the $xz/yz$ orbital is more coherent than
the $xy$ orbital.
The scattering rate of the $xz/yz$ orbital is
nearly proportional to $T^{2}$ from 0 K (0 K$^{2}$)
to 265 K (70000 K$^{2}$), implying that
a Fermi liquid behaviour for this temperature range.
The scattering rate of the $xy$ orbital is nearly
proportional to $T^{2}$ from 0 K (0 K$^{2}$)
to 100 K (10000 K$^{2}$), and
there is a coherence-incoherence crossover
around the temperature of 150 K.
This result of the scattering rate (Fig.\ref{fig:tau})
from the transport data of Fig.\ref{fig:sigma}
implies that
(i) the $xy$ orbital is more incoherent than the $xz/yz$ orbital, and
(ii) there is no anomalous behaviour for the $xz/yz$ orbital.

To check the validity of the orbital dependent
spatially local scattering rate in Fig.\ref{fig:tau},
we compute magnetoresistance (MR) coefficient, $\alpha$,
from the scattering rate using the formulation of Eq.\ref{eq:alpha}.
Fig.\ref{fig:alpha} compares the computed
MR coefficient with its experimental value of Ref.\cite{rullier2012multiorbital}.
It is shown that the computed MR coefficient is
consistent with the experimental MR coefficient,
validates the local ansatz for the orbital dependent scattering
rate.

\begin{eqnarray} \label{eq:alpha}
&\alpha(T)=&\frac{\delta\rho(T,H)}{\rho(T,0)H^{2}} \nonumber\\
&\frac{\delta\rho(T,H)}{\rho(T,0)} =& -\frac{\delta\sigma(T,H)}{\sigma(T,0)} - (\frac{\sigma_{xy}}{\sigma_{xx}})^{2} \nonumber\\
&\sigma_{xy}=&H\sigma_{xx}^{2}R_{H} \nonumber\\
&(\frac{\sigma_{xy}}{\sigma_{xx}})^{2} =& H^{2}\sigma_{xx}^{2}R_{H}^{2}=H^{2}\sigma^{2}R_{H}^{2} \nonumber\\
&-\frac{\delta\sigma(T,H)}{\sigma(T,0)}=&[(\sum_{\nu}n_{\nu}e\mu_{\nu}^{3})/\sigma ]H^{2}
\end{eqnarray}

\bibliography{refs_LiFeAs_Nonlocal}

\end{document}